\documentclass[twocolumn,english]{IEEEtran}
\usepackage[T1]{fontenc}
\usepackage{color}
\usepackage{babel}
\usepackage{array}
\usepackage{amsmath}
\usepackage{amsthm}
\usepackage{amssymb}
\usepackage{stackrel}
\usepackage{graphicx}
\usepackage[unicode=true,
 bookmarks=true,bookmarksnumbered=true,bookmarksopen=true,bookmarksopenlevel=1,
 breaklinks=false,pdfborder={0 0 0},pdfborderstyle={},backref=false,colorlinks=false]
 {hyperref}
\hypersetup{pdftitle={Your Title},
 pdfauthor={Your Name},
 pdfpagelayout=OneColumn, pdfnewwindow=true, pdfstartview=XYZ, plainpages=false}

\makeatletter

\providecommand{\tabularnewline}{\\}

\theoremstyle{plain}
\newtheorem{thm}{\protect\theoremname}
\theoremstyle{plain}
\newtheorem{prop}[thm]{\protect\propositionname}

\usepackage{balance}
\usepackage{color}
\usepackage[linesnumbered,lined,ruled,commentsnumbered,noend]{algorithm2e}
\usepackage{ragged2e}
\newcommand\justfy{\noindent\justifying}
\newtheorem{property}{Property}
\makeatother

\providecommand{\propositionname}{Proposition}
\providecommand{\theoremname}{Theorem}

\begin{document}
\title{Optimal Spectrum Partitioning and Licensing in Tiered Access under
Stochastic Market Models}
\author{Gourav~Saha,~and Alhussein~A.~Abouzeid\thanks{G. Saha and A. A. Abouzeid are with the Department of Electrical,
Computer and Systems Engineering, Rensselaer Polytechnic Institute,
Troy, NY 12180, USA; Email: sahag@rpi.edu, abouzeid@ecse.rpi.edu.
A preliminary version of this paper has been accepted for publication
in WiOpt 2020.}}
\maketitle
\begin{abstract}
We consider the problem of partitioning a spectrum band into $M$
channels of equal bandwidth, and then further assigning these M channels
into $P$ licensed channels and $M-P$ unlicensed channels. Licensed
channels can be accessed both for licensed and opportunistic use following
a tiered structure which has a higher priority for licensed use. Unlicensed
channels can be accessed only for opportunistic use. We address the
following question in this paper. Given a market setup, what values
of $M$ and $P$ maximize the net spectrum utilization of the spectrum
band? While this problem is of fundamental nature, it is highly relevant
practically, e.g., in the context of partitioning the recently proposed
Citizens Broadband Radio Service band. If $M$ is too high or too
low, it may decrease spectrum utilization due to limited channel capacity
or due to wastage of channel capacity, respectively. If $P$ is too
high (low), it will not incentivize the wireless operators who are
primarily interested in unlicensed channels (licensed channels) to
join the market. These tradeoffs are captured in our optimization
problem which manifests itself as a two-stage Stackelberg game. We
design an algorithm to solve the Stackelberg game and hence find the
optimal $M$ and $P$. The algorithm design also involves an efficient
Monte Carlo integrator to evaluate the expected value of the involved
random variables like spectrum utilization and operators' revenue.
We also benchmark our algorithms using numerical simulations.
\end{abstract}

\begin{IEEEkeywords}
Spectrum License, Opportunistic Spectrum Access, CBRS band, Stackelberg
game, Iterated Removal of Strictly Dominated Strategies, Monte Carlo
Integration, Optimization\vspace{-1.0em}
\end{IEEEkeywords}

\section{Introduction\label{sec:Introduction}}

To support the ever growing wireless data traffic, the Federal Communication
Commission (FCC) released the underutilized Citizens Broadband Radio
Service (CBRS) band for shared use in 2015 \cite{fcc2015}. CBRS band
is a $150\:MHz$ federal spectrum band from $3.55\:GHz$ to $3.7\:GHz$.
The $150\:MHz$ band is divided into $15$ channels of $10\:MHz$
each. The shared use of the CBRS band follows an order of priority.
Federal users have the highest priority access to the channels. Out
of the $15$ channels, $7$ are Priority Access Licenses (PALs). PAL
licenses are sold through auctions and the lease duration of a PAL
license may range between $1-10$ years \cite{fcc2015,fcc2016,fcc2017}.
A PAL license holder can use their channel only if federal users are
not using it. The remaining $8$ out of the $15$ channels are reserved
only for opportunistic use by General Authorized Access (GAA) users.
Opportunistic channel allocation to GAA users can happen at a time
scale of minutes to weeks. GAA users can use these $8$ channels if
federal users are not using the channels. GAA users can also use the
$7$ PAL channels provided that neither federal users nor PAL license
holders are using it.

\begin{figure}[t]
\begin{centering}
\includegraphics[scale=0.34]{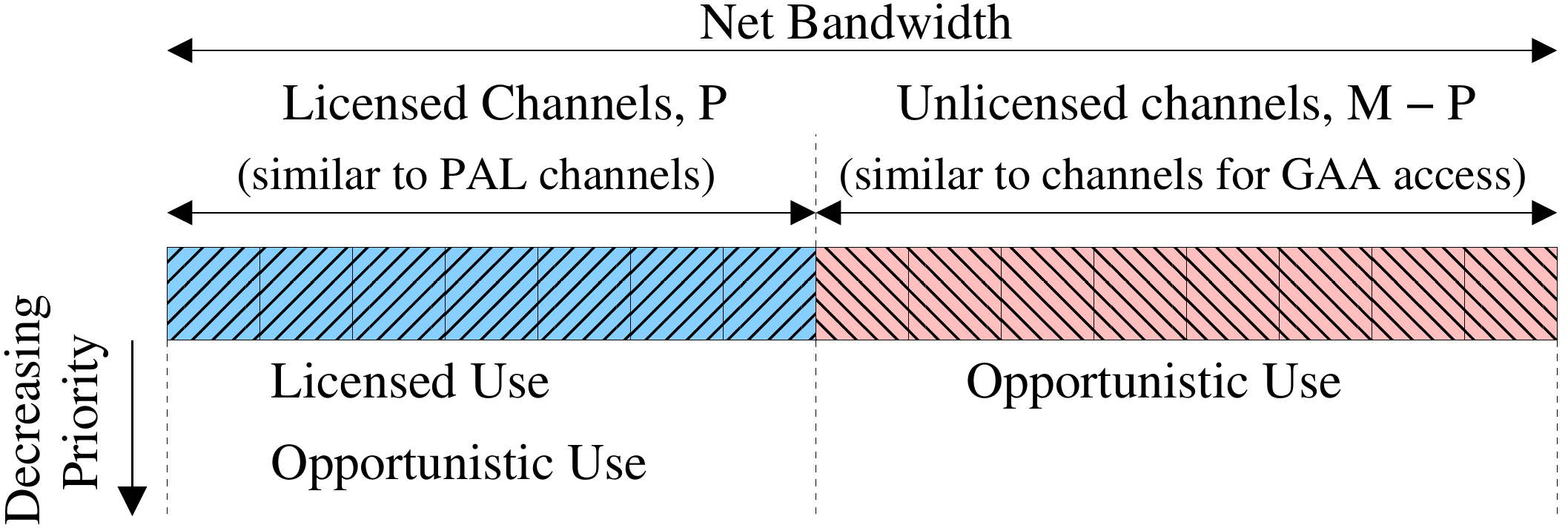}
\par\end{centering}
\caption{Pictorial representation of the tiered spectrum model under consideration
in this paper.\vspace{-3.0em}}
\label{cbrs_partition}
\end{figure}

As mentioned in the previous paragraph, the CBRS band is divided into
$M=15$ channels out of which there are $P=7$ PAL licenses. But does
$M=15$ and $P=7$ maximize the utilization of the CBRS band? In this
paper, we are interested in the following abstraction of this question
whose application is not limited to CBRS band. A net bandwidth is
partitioned into $M$ channels of equal bandwidth. These $M$ channels
are further divided into $P$ licensed channels (similar to PAL channels)
and $M-P$ unlicensed channels (similar to channels reserved for GAA
users). In this paper, the process of dividing the net bandwidth into
$M$ channels is called \textit{spectrum partitioning} and the process
of allocating these $M$ channels as licensed and unlicensed channels
is called \textit{spectrum licensing}.\textcolor{blue}{{} }Licensed
channels are used for both licensed use and opportunistic use with
the former having higher priority. Unlicensed channels are reserved
for opportunistic use only. This spectrum access model is shown in
Figure \ref{cbrs_partition}. The wireless operators earn revenue
by serving customer demands. A wireless operator is incentivized to
join the market if the revenue which it can earn is above a desired
threshold. For the given setup, what value of $M$ and $P$ maximizes
spectrum utilization where spectrum utilization is defined as the
net amount of customer demand served by the entire bandwidth?

\noindent 

\vspace{-1.0em}

There are various factors that decide the optimal values of $M$ and
$P$. Some of these factors are as follows. If the number of channels
$M$ increases, the bandwidth, and hence capacity, of each channel
decreases. The capacity of each channel should be large enough to
accommodate a good portion of the customer demand of a wireless operator
but not so large that most of the capacity of the channel is not utilized
for majority of the time. This suggests that $M$ should not be too
small or too large. If the number of licensed channels $P$ is too
high, there is a small number of unlicensed channels. Therefore, those
operators who primarily rely on unlicensed channels to serve customer
demands will not be able to generate enough revenue and hence will
not be incentivized to join the market. Similarly, if the $P$ is
too low, wireless operators who primarily rely on licensed channels
to serve customer demands will not be incentivized to join the market.
$P$ should be set such that enough operators join the market to ensure
that the customer demands served over the entire bandwidth is as high
as possible. There may be other qualitative factors governing optimal
$M$ and $P$. Therefore, in this paper, we design an algorithm to
jointly optimize $M$ and $P$ such that spectrum utilization is maximized.\vspace{-1.0em}

\subsection{Related Work\label{subsec:Related-Work}}

Variations of the spectrum partitioning and spectrum licensing problems
considered in this paper have been studied separately, but not jointly,
in the spectrum sharing and related fields. There are a few works
that have addressed problems similar to partitioning of a fixed bandwidth
into optimal number of channels. In \cite{partition1_andrews}, the
authors derive an analytical expression for the optimal number of
channels such that the spatial density of transmission is maximized
subject to a fixed link transmission rate and packet error rate. Partitioning
of bandwidth in the presence of guard bands has been considered in
\cite{flexauc} where the authors used Stackelberg game formulation
to analyze how a spectrum holder should partition its bandwidth in
order to maximize its revenue in spectrum auctions.

The second problem studied in this paper essentially deals with spectrum
licensing. This has been widely studied in the literature from various
perspectives. Some works concentrated on minimizing the amount of
bandwidth allocated to backup channels (unlicensed channels in our
case) while providing a certain level of guarantee to secondary users
against channel preemption \cite{reliability}. There has also been
research on overlay D2D and cellular devices which studied optimal
partitioning of orthogonal in-band spectrum to maximize the average
throughout rates of cellular and D2D devices \cite{d2doptimal}. In
\cite{bazelon}, the authors investigated whether to allocate an additional
spectrum band for licensed or unlicensed use and concluded that the
licensed use is more favourable for maximizing the social surplus.
A similar result has been shown in \cite{berry_unlicensed} which
studied the effect of adding an unlicensed spectrum band in a market
consisting of wireless operators with licensed channels. The authors
showed that if the amount of unlicensed spectrum band is below a certain
limit, the the overall social welfare may decrease with the increase
in unlicensed spectrum band. The authors in \cite{ghosh2019competition,ghosh2020entry}
studied the CBRS band for a market setup that consists of Environmental
Sensing Capability operators (ESCs) whose sole job is to monitor and
report spectrum occupancy to the wireless operators. The authors analyzed
how the ratio of the licensed and unlicensed bands affects the market
competition between the ESC operators, the wireless operators, and
the end users of the CBRS band. There is a line of work which studies
spectrum partitioning for topics similar to licensed and unlicensed
use using Stackelberg games; macro cells and small cells \cite{berry_stackelberg,hossain_stackelberg},
long-term leasing market and short-term rental market \cite{head},
and 4G cellular and Super Wifi services \cite{constantleaseprice}.

Such a diverse body of work just on spectrum partitioning and licensing
is justified because individual problem setups have their own salient
features and hence require their own analysis. Our problem setup considers
\textit{jointly} optimizing spectrum partitioning and spectrum licensing,
which has not been considered in the existing literature. This problem
is novel because of the combination of the following two reasons.
\textit{First}, our spectrum access model, like CBRS, is a combination
of (a) Unlicensed spectrum access model. This is because $M-P$ unlicensed
channels are reserved specifically for opportunistic use. (b) Primary-secondary
spectrum access model. This is because $P$ licensed channels can
be used for opportunistic access following the priority hierarchy.
Prior works like \cite{berry_unlicensed,ghosh2019competition,ghosh2020entry},
which solved the spectrum licensing problem, did not simultaneously
consider both of the spectrum access models. \textit{Second}, we consider
a very generalized system model in terms of the number of operators,
their types, and their heterogeneity. Such a setup leads to a scenario
where the regulator has to decide $M$ and $P$ such that the right
set of wireless operators are incentivized to join the market.\vspace{-1.0em}

\subsection{Contribution and Paper Organization\label{subsec:Contribution-and-Paper}}

We now present an overall outline of the paper and, in the process,
discuss its main contributions. In Section \ref{sec:System-Model},
we present a system model which can mathematically capture the effect
of the number of channels, $M$, and number of licensed channels,
$P$, on the spectrum utilization. The proposed system model captures
spectrum auctions using a simple stochastic model without going into
complex game-theoretic formulations. Based on our reading of \cite{fcc2015,fcc2016,fcc2017}
and other literature on CBRS band, it is not clear if there is a consensus
in the literature/policy about whether PAL license holders are also
allowed to use channels opportunistically. So it is possible that
PAL license holders may or may not be allowed to use channels opportunistically.
Our model is general enough to capture both of these cases. Our model
also generalizes well to various opportunistic access strategies like
overlay or interweave spectrum access \cite{survey1}.

It is possible that a choice of values of $M$ and $P$ that incentivizes
one group of wireless operators may not incentivize another group.
Therefore, a $M$ and $P$ which incentivizes all the wireless operators
may not exist. This argument can be exemplified by refering to \cite{fcc2015,fcc2016,fcc2017}
which shows a lot of debate between the wireless operators concerning
the parameters of the CBRS model. Even if it is possible to satisfy
all the operators, it may not be optimal to do so in terms of maximizing
the spectrum utilization. We capture this idea by formulating our
problem as a two-stage Stackelberg game in Section \ref{subsec:Stackelberg}
which forms the second contribution of the paper. The Stackelberg
game consists of the regulator (leader) and the wireless operators
(followers). In Stage 1, the regulator sets $M$ and $P$ to maximize
spectrum utilization. In Stage 2, the wireless operators decide whether
or not to join the market based on the $M$ and $P$ set by the regulator
in Stage 1.

In Section \ref{subsec:Solution}, we design an algorithm to solve
the Stackelberg game and hence the optimal $M$ and $P$ which maximize
spectrum utilization. We approach this in steps. Few properties associated
with expected revenue of an operator are discussed first. We show
that when these properties hold, we can design a polynomial time algorithm
to solve Stage 2 of the Stackelberg game. We finally solve Stage 1
of the Stackelberg game using a grid search approach to find the optimal
$M$ and $P$ which maximizes spectrum utilization. \textit{To the
best of our knowledge, joint optimization of partitioning and tiered
licensing have not been considered in the existing related literature.}
Hence, designing an algorithm for joint optimization of $M$ and $P$
is the fourth contribution of the paper.

The solve the Stackelberg game, we have to calculate the expected
revenue of an operator and expected spectrum utilization. The complex
nature of the problem does not allow simple analytical formulas of
these expected values. Even if such analytical formulas are possible,
adapting them to changes in system model can be time consuming. Therefore,
we develop a Monte Carlo integrator to evaluate these expected values
in Section \ref{sec:Monte-Carlo-Integrator}. Our choice of using
a Monte Carlo integrator over deterministic numerical integration
techniques is because our setup involves evaluation of high-dimensional
integrals. Unlike deterministic numerical integration techniques,
the computation time of Monte carlo integration does not scale with
dimension. One of the main bottlenecks of Monte Carlo integration
is random sampling. While designing our Monte Carlo integrator, we
reduced random sampling as much as possible to make it more time efficient.
Designing an efficient Monte Carlo integrator which can easily adapt
to few changes in the system model is the third contribution of the
paper.

Finally, we use the algorithms designed in Sections \ref{subsec:Solution}
and \ref{sec:Monte-Carlo-Integrator} to obtain important numerical
results in Section \ref{sec:Numerical-Results} which constitutes
the final contribution of the paper. Our numerical results show how
optimal values of $M$ and $P$ vary with market parameters.\vspace{-0.5em}

\section{System Model\label{sec:System-Model}}

In this section, we discuss individual components of our system model
in Sections \ref{subsec:Channel-Model} to \ref{subsec:Allocation-Model}.
The list of important notations is included in Table \ref{notations}.
We introduce three set theoretic notations. Consider two sets $\mathcal{A}$
and $\mathcal{B}$. The operation $\mathcal{A}\bigcup\mathcal{B}$
implies the union $\mathcal{A}$ and $\mathcal{B}$. The operation
$\mathcal{A}\backslash\mathcal{B}$ is as set which consists of all
those elements in $\mathcal{A}$ which are not in $\mathcal{B}$.
A singleton set consisting of element $a$ is denoted by $\left\{ a\right\} $.

\noindent 
\begin{table}[t]
\caption{A table of important notations. \label{notations}}

\begin{centering}
\begin{tabular}{|>{\centering}m{1.1cm}|m{6.7cm}|}
\hline 
\textbf{Notation} & \textbf{Description}\tabularnewline
\hline 
$t$ , $\gamma$ & $t^{th}$ time slot and epoch $\gamma$ resp.\tabularnewline
\hline 
$M$ , $P$ & Number of channels and number of licensed channels resp.\tabularnewline
\hline 
$\phi$ & Indicator variable which decides if a Tier-1 operator can also use
channels opportunistically. We have, $\phi\in\left\{ 0,1\right\} $.\tabularnewline
\hline 
$D$ & Capacity of the entire bandwidth for licensed access.\tabularnewline
\hline 
$\alpha_{L}$ , $\alpha_{U}$ & Interference parameter associated with opportunistic use of licensed
and unlicensed channels resp.\tabularnewline
\hline 
$\mathcal{S}_{L}^{C}$ , $\mathcal{S}_{U}^{C}$ & Set of candidate licensed and unlicensed operators resp.\tabularnewline
\hline 
$\mathcal{S}_{L}$ , $\mathcal{S}_{U}$ & Set of interested licensed and unlicensed operators resp.\tabularnewline
\hline 
$\mathcal{S}$ & Set of interested operators; $\mathcal{S}=\mathcal{S}_{L}\bigcup\mathcal{S}_{U}$.\tabularnewline
\hline 
$\mathcal{T}_{1}\left(\gamma\right)$ & Set of Tier-1 operators in epoch $\gamma$, i.e. set of interested
licensed operators who won licensed channels in epoch $\gamma$.\tabularnewline
\hline 
$\mathcal{T}_{2}\left(\gamma\right)$ & Set of Tier-2 operators in epoch $\gamma$; $\mathcal{T}_{2}\left(\gamma\right)=\mathcal{S}\backslash\mathcal{T}_{1}\left(\gamma\right)$.\tabularnewline
\hline 
$x_{k}\left(t\right)$ & Customer demand of the $k^{th}$ operator in $t^{th}$ time slot.\tabularnewline
\hline 
$\mu_{k}^{\theta}$ , $\sigma_{k}^{\theta}$ & Mean and standard deviation resp. of Gaussian random variable $\theta_{k}\left(t\right)$
where, $x_{k}\left(t\right)=\max\left(0,\theta_{k}\left(t\right)\right)$.\tabularnewline
\hline 
$\widetilde{x}_{k,i}\left(t\right)$ & Amount of demand served by the $k^{th}$ operator in $t^{th}$ time
slot if it is a Tier-$i$ operator where $i\in\left\{ 1,2\right\} $.\tabularnewline
\hline 
$\widetilde{x}_{k,a}\left(t\right)$ & Amount of demand served by the $k^{th}$ operator in $t^{th}$ time
slot when spectrum access type is $a$ where $a\in\left\{ lc,op\right\} $.\tabularnewline
\hline 
$X_{k,a}\left(\gamma\right)$ & Net demand served by the $k^{th}$ operator in epoch $\gamma$ when
spectrum access type is $a$ where $a\in\left\{ lc,op\right\} $.\tabularnewline
\hline 
$R_{k,a}\left(\gamma\right)$ & Revenue of $k^{th}$ operator in epoch $\gamma$ using spectrum access
of type $a$ where $a\in\left\{ lc,op\right\} $.\tabularnewline
\hline 
$R_{k,i}\left(\gamma\right)$ & Revenue of $k^{th}$ operator in epoch $\gamma$ if it is a Tier-$i$
operator.\tabularnewline
\hline 
$V_{k}\left(\gamma\right)$ & Bid of the $k^{th}$ operator, where $k\in\mathcal{S}_{L}$, in epoch
$\gamma$.\tabularnewline
\hline 
$h_{k}\left(\cdot\right)$ & A function associated with $k^{th}$ operator which maps the mean
of $X_{k,a}\left(\gamma\right)$ to the mean of $R_{k,a}\left(\gamma\right)$.\tabularnewline
\hline 
$\mu_{k,a}^{X}\,,\,\sigma_{k,a}^{X}$ & Mean and standard deviation of $X_{k,a}\left(\gamma\right)$.\tabularnewline
\hline 
$\mu_{k,a}^{R}\,,\,\sigma_{k,a}^{R}$ & Mean and standard deviation of $R_{k,a}\left(\gamma\right)$. We have,
$\mbox{{\ensuremath{\mu_{k,a}^{R}}=\ensuremath{h_{k}\left(\mu_{k,a}^{X}\right)}}}$.\tabularnewline
\hline 
$\rho_{k}$ & \noindent Correlation coefficent between $X_{k,a}\left(\gamma\right)$
and $R_{k,a}\left(\gamma\right)$.\tabularnewline
\hline 
$\omega_{k}$ & \noindent Correlation coefficent between $V_{k}\left(\gamma\right)$
and $R_{k,lc}\left(\gamma\right)$.\tabularnewline
\hline 
$\lambda_{k}$ & Minimum revenue requirement of the $k^{th}$ operator.\tabularnewline
\hline 
$\xi_{k}$ & The tuple $\left(\mu_{k}^{\theta},\,\sigma_{k}^{\theta},\,h_{k}\left(\cdot\right),\,\sigma_{k,a}^{R},\,\rho_{k},\,\rho_{k},\,\lambda_{k}\right)$
associated with the $k^{th}$ operator.\tabularnewline
\hline 
$\left(x\right)^{+}$ & $\left(x\right)^{+}=\max\left(0,x\right)$\tabularnewline
\hline 
\end{tabular}
\par\end{centering}
\vspace{-1.0em}
\end{table}

\vspace{-1.5em}

\subsection{Channel Model\label{subsec:Channel-Model}}

A net bandwidth of $W\:hz$ is divided into $M$ channels of equal
bandwidth $\frac{W}{M}$. Out of the $M$ channels, $P$ channels
are licensed channels while the remaining $M-P$ channels are unlicensed
channels. In our model, time is divided into slots where $t\in\mathbb{Z}^{+}$
denotes the $t^{th}$ time slot. Licensed channels are allocated for
prioritized licensed use and opportunistic use while unlicensed channels
are allocated only for opportunistic use. Allocation of licensed channels
for licensed use happens through auctions. These auctions occur every
$T\geq$1 time slots where $T$ is the lease duration. An entire lease
duration is called an ``epoch''. Epoch $\gamma$ is from time slot
$\left(\gamma-1\right)T+1$ to $\gamma T$. Allocation of licensed
channels and unlicensed channels for opportunistic use occur every
time slot.

Those operators who are allocated licensed channels for licensed use
are called \textit{Tier-1 operators} while those who are not are called
\textit{Tier-2 operators}. In our model, an operator can be allocated
atmost one licensed channel for licensed use in an epoch, i.e. spectrum
cap is one. Similar assumptions has been made in prior works like
\cite{SpectrumCap1}. Spectrum cap of one ensures fairness by allocating
the licensed channels to as many operators as possible. A Tier-1 operator
can also use opportunistic channels to serve its customer demand in
case the bandwidth of the allocated licensed channel is not sufficient.
Tier-2 operators uses channels opportunsitically. Tier-1 operators
may also use channels opportunistically. Let $\phi\in\left\{ 0,1\right\} $
be an indicator variable which decides if Tier-1 operators can use
channels opportunistically. If $\phi=1$, then Tier-1 operators can
use channels opportunistically and otherwise they cannot.

The capacity of a channel/bandwidth is the maximum units of customer
demand that can be served using that channel/bandwidth in a time slot.
Let $D$ be the capacity of the entire bandwidth of $W\:hz$ when
used for licensed access. As the entire bandwidth is partitioned into
$M$ channels, each licensed channel has a capacity of $\frac{D}{M}$
when used for licensed use while the unlicensed channels has a capacity
of $\frac{\alpha_{U}D}{M}$ where $\alpha_{U}\in\left[0,1\right]$
is the interference parameter associated with unlicensed channels
for opportunistic use. Licensed channels can also be used for opportunistic
use following the priority hierarchy shown in Figure \ref{cbrs_partition}.
Let the customer demand of a Tier-1 operator be $d$ units. It will
use its licensed channel to serve its customer demand. The remaining
capacity of the licensed channel which can be utilized for opportunsitic
use is given by the function $\mathcal{C}\left(d,\alpha_{L}\right)$
where $\alpha_{L}\in\left[0,1\right]$ is the interference parameter
associated with licensed channels for opportunistic use. The expression
for $\mathcal{C}\left(d,\alpha_{L}\right)$ depends on the opportunistic
spectrum access (OSA) strategy: overlay or interweave \cite{survey1}.
For \textit{overlay spectrum access}, $\mathcal{C}\left(d,\alpha_{L}\right)=\alpha_{L}\left(\frac{D}{M}-d\right)^{+}$,
where $\left(x\right)^{+}=\max\left(0,x\right)$. For \textit{interweave
spectrum access}, $\mathcal{C}\left(d,\alpha_{L}\right)$ is equal
to $0$ if $d>0$ and equal to $\frac{\alpha_{L}D}{M}$ if $d=0$.
Parameters $\alpha_{L}$ and $\alpha_{U}$ capture the lower efficiency
of opportunistic use as compared to licensed use \cite{fcc2015}.
In general, we expect $\alpha_{L}\leq\alpha_{U}$. This may happen
because the transmission power cap for opportunistic use maybe lower
for licensed channels compared to unlicensed channels in order to
protect Tier-1 operators from harmful interference.\vspace{-0.75em}

\subsection{Operators, their Demand and Revenue Model\label{subsec:Demand-Model}}

The market consists of the \textit{candidate licensed operators} denoted
by $\mathcal{S}_{L}^{C}$ and the \textit{candidate unlicensed operators}
denoted by $\mathcal{S}_{U}^{C}$ where $\mathcal{S}_{L}^{C}$ and
$\mathcal{S}_{U}^{C}$ are disjoint sets. A candidate licensed operator
is a Tier-1 operator in those epochs in which it is allocated a licensed
channel in the auction and a Tier-2 operator in those epochs in which
it is not allocated a licensed channel. A candidate unlicensed operator
is always a Tier-2 operator. A candidate operator has to invest in
infrastructure development if it wants to join the market.

All the candidate operators have to invest in infrastructure development
to join the market. In order to generate return on infrastructure
cost and the cost of leasing a channel, a candidate operator wants
to earn a minimum expected revenue in an epoch. Let $\lambda_{k}$
be the minimum expected revenue (MER) of the $k^{th}$ operator. A
subset of candidate licensed and unlicensed operators are interested
in joining the market if the value of $M$ and $P$ set by the regulator
is such that the expected revenue of the operator in an epoch is greater
than its MER. The set of \textit{interested licensed operators} and
\textit{interested unlicensed operators} are denoted by $\mathcal{S}_{L}$
and $\mathcal{S}_{U}$ respectively. We have $\mathcal{S}_{L}\subseteq\mathcal{S}_{L}^{C}$
and $\mathcal{S}_{U}\subseteq\mathcal{S}_{U}^{C}$. The set of operators,
$\left(\mathcal{S}_{L}^{C}-\mathcal{S}_{L}\right)\bigcup\left(\mathcal{S}_{U}^{C}-\mathcal{S}_{U}\right)$,
does not join the market. A candidate licensed/unlicensed operator
gets to decide whether to join or not join the market only once. An
operator gets to participate in auctions for licensed channels or
to use channels opportunistically only if it decides join the market.

In our model, every operator has a separate pool of customers each
with its own stochastic demands, i.e. we do not model price competition
between operators to attract a common pool of customers. Consider
the $t^{th}$ time slot of epoch $\gamma$. The customer demand, or
simply demand, of the $k^{th}$ operator in the $t^{th}$ time slot
is $x_{k}\left(t\right)$. In our model, $x_{k}\left(t\right)=\max\left(0\,,\,\theta_{k}\left(t\right)\right)$
where $\theta_{k}\left(t\right)$ are iid Gaussian random variable\footnote{All the iid random variables used throughout the paper are identical
with respect to time slot, $t$, or epoch, $\gamma$, and \textit{not}
with respect to operator index $k$.} with mean $\mu_{k}^{\theta}$ and standard deviation $\sigma_{k}^{\theta}$,
i.e. $\theta_{k}\left(t\right)\sim\mathcal{N}\left(\mu_{k}^{\theta},\left(\sigma_{k}^{\theta}\right)^{2}\right)\,,\,\forall t$.
The $k^{th}$ operator may be able to serve only a fraction of the
customer demand. Let $\widetilde{x}_{k,1}\left(t\right)$ and $\widetilde{x}_{k,2}\left(t\right)$
denote the amount of customer demand served by the $k^{th}$ operator
if it is a Tier-1 and a Tier-2 operator respectively in epoch $\gamma$.
We have $\widetilde{x}_{k,1}\left(t\right)\leq x_{k}\left(t\right)$
and $\widetilde{x}_{k,2}\left(t\right)\leq x_{k}\left(t\right)$.
$\widetilde{x}_{k,1}\left(t\right)$ and $\widetilde{x}_{k,2}\left(t\right)$
can be expressed as follows
\begin{eqnarray}
\widetilde{x}_{k,1}\left(t\right) & = & \widetilde{x}_{k,lc}\left(t\right)+\widetilde{x}_{k,op}\left(t\right)\label{eq:2.2.a}\\
\widetilde{x}_{k,2}\left(t\right) & = & \widetilde{x}_{k,op}\left(t\right)\label{eq:2.2.b}
\end{eqnarray}
where\textcolor{blue}{{} }$\widetilde{x}_{k,lc}\left(t\right)=\min\left(x_{k}\left(t\right),\frac{D}{M}\right)$.
The term $\widetilde{x}_{k,lc}\left(t\right)$ in (\ref{eq:2.2.a})
is the amount of customer demand served a Tier-1 operator using the
channel allocated to it for licensed use. The term $\widetilde{x}_{k,op}\left(t\right)$
in (\ref{eq:2.2.a}) and (\ref{eq:2.2.b}) is the demand served by
an operator by using channels opportunistically. It will be shown
in Section \ref{subsec:Allocation-Model} that $\widetilde{x}_{k,op}\left(t\right)$
is a iid random variable. Also, if $\phi=0$, then a Tier-1 operator
cannot use channels opportunistically and hence $\widetilde{x}_{k,op}\left(t\right)\equiv0$
in (\ref{eq:2.2.a}). In (\ref{eq:2.2.a}) and (\ref{eq:2.2.b}),
$\widetilde{x}_{k,1}\left(t\right)$ and $\widetilde{x}_{k,2}\left(t\right)$
can be expressed as a time invariant function of iid random variables
$x_{k}\left(t\right)$ and $\widetilde{x}_{k,op}\left(t\right)$.
Therefore, $\widetilde{x}_{k,1}\left(t\right)$ and $\widetilde{x}_{k,2}\left(t\right)$
are iid random variables as well. Let $\widetilde{\mu}_{k,a}^{x}$
and $\widetilde{\sigma}_{k,a}^{x}$ denote the mean and standard deviation
of $\widetilde{x}_{k,a}\left(t\right)$ respectively. We have,\vspace{-1.0em}

\begin{equation}
\widetilde{\mu}_{k,lc}^{x}=\stackrel[0]{\frac{D}{M}}{\int}\vartheta f_{k}^{\theta}\left(\vartheta\right)\,d\vartheta+\frac{D}{M}\stackrel[\frac{D}{M}]{\infty}{\int}f_{k}^{\theta}\left(\vartheta\right)\,d\vartheta\label{eq:2.2.0.1}
\end{equation}

\vspace{-1.0em}

\begin{equation}
\left(\widetilde{\sigma}_{k,lc}^{x}\right)^{2}=\stackrel[0]{\frac{D}{M}}{\int}\vartheta^{2}f_{k}^{\theta}\left(\vartheta\right)\,d\vartheta+\left(\frac{D}{M}\right)^{2}\stackrel[\frac{D}{M}]{\infty}{\int}f_{k}^{\theta}\left(\vartheta\right)\,d\vartheta-\left(\widetilde{\mu}_{k,lc}^{x}\right)^{2}\label{eq:2.2.0.2}
\end{equation}

\noindent where $f_{k}^{\theta}\left(\vartheta\right)$ is the probability
density function of $\theta_{k}\left(t\right)$. In general, an analytical
expression for $\widetilde{\mu}_{k,op}^{x}$ and $\widetilde{\sigma}_{k,op}^{x}$
is not possible because of the complex nature of opportunistic spectrum
allocation algorithm. We have designed a Monte Carlo integrator which
can compute $\widetilde{\mu}_{k,op}^{x}$ in Section \ref{sec:Monte-Carlo-Integrator}.

Throughout the rest of the paper we will use the subscript $k,i$,
where $i\in\left\{ 1,2\right\} $, to denote variables associated
with $k^{th}$ operator when its is a Tier-$i$ operator. Also, we
will use the subscript $k,a$, where $a\in\left\{ lc,op\right\} $,
to denote variables associated with $k^{th}$ operator when access
type is licensed ($a=lc$) or opportunistic ($a=op$). Let $X_{k,a}\left(\gamma\right)$
denote the net demand served by the $k^{th}$ operator in epoch $\gamma$
when access type is $a$. Mathematically,\vspace{-0.5em}

\begin{equation}
X_{k,a}\left(\gamma\right)=\stackrel[t=\left(\gamma-1\right)T+1]{\gamma T}{\sum}\widetilde{x}_{k,a}\left(t\right)\;;\:a\in\left\{ lc,op\right\} \label{eq:2.2.}
\end{equation}

Since $\widetilde{x}_{k,a}\left(t\right)$ is iid random variable
and the lease duration $T$ is quite large in practice, $X_{k,a}\left(\gamma\right)$
can be approximated as a Gaussian random variable using \textit{Central
Limit Theorem}\textcolor{blue}{{} }\cite[Chapter 8]{clt_ross}. The
mean, $\mu_{k,a}^{X}$, and standard deviation, $\sigma_{k,a}^{X}$,
of $X_{k,a}\left(\gamma\right)$ are given by 
\begin{equation}
\mu_{k,a}^{X}=\widetilde{\mu}_{k,a}^{x}T\qquad;\qquad\sigma_{k,a}^{X}=\widetilde{\sigma}_{k,a}^{x}\sqrt{T}\label{eq:2.2.2}
\end{equation}

To this end we have, $X_{k,a}\left(\gamma\right)\sim\mathcal{N}\left(\mu_{k,a}^{X},\left(\sigma_{k,a}^{X}\right)^{2}\right)\,,\,\forall\gamma$.

\textit{Remark 1: Gaussian nature of }$X_{k,a}\left(\gamma\right)$\textit{.}
$X_{k,a}\left(\gamma\right)$ is always a positive quantity because
net demand served is always positive. But, we approximated $X_{k,a}\left(\gamma\right)$
as a Gaussian random variable and hence the approximated $X_{k,a}\left(\gamma\right)$
can be negative. However, the probability of $X_{k,a}\left(\gamma\right)$
being negative is\vspace{-0.5em}
\[
P\left[X_{k,a}\left(\gamma\right)<0\right]=\frac{1}{2}\left(1+\text{erf}\left(-\frac{\widetilde{\mu}_{k,a}^{x}\sqrt{T}}{\sqrt{2}\widetilde{\sigma}_{k,a}^{x}}\right)\right)
\]
where $\text{erf}\left(\cdot\right)$ is the error function. For all
practical setup, $T$ is large enough that $P\left[X_{k,a}\left(\gamma\right)<0\right]$
is very small. The use of Gaussian model for non-negative random variables
have been used in prior works like \cite{gaussian1}.

An operator generates revenue by serving customer demand. Let $R_{k,a}\left(\gamma\right)$
denote the revenue earned by the $k^{th}$ operator in epoch $\gamma$
when access type is $a$. We model $R_{k,a}\left(\gamma\right)$ as
a random variable which follows the stochastic model
\begin{equation}
\begin{bmatrix}X_{k,a}\left(\gamma\right)\\
R_{k,a}\left(\gamma\right)
\end{bmatrix}\sim\mathcal{N}\left(\begin{bmatrix}\mu_{k,a}^{X}\\
\mu_{k,a}^{R}
\end{bmatrix},\begin{bmatrix}\left(\sigma_{k,a}^{X}\right)^{2} & \rho_{k}\sigma_{k,a}^{X}\sigma_{k,a}^{R}\\
\rho_{k}\sigma_{k,a}^{X}\sigma_{k,a}^{R} & \left(\sigma_{k,a}^{R}\right)^{2}
\end{bmatrix}\right)\label{eq:2.2.3}
\end{equation}

\noindent for all $\gamma$ where $\mu_{k,a}^{R}=h_{k}\left(\mu_{k,a}^{X}\right)$.
According to (\ref{eq:2.2.3}), the net demand served and the net
revenue earned in epoch $\gamma$ are jointly Gaussian. The mean of
$R_{k,a}\left(\gamma\right)$ is $\mu_{k,a}^{R}=h_{k}\left(\mu_{k,a}^{X}\right)$
where $h_{k}\left(\mu_{k,a}^{X}\right)$ is a monotonic increasing
function of the mean demand served by the $k^{th}$ operator in an
epoch, $\mu_{k,a}^{X}$. The standard deviation of $R_{k,a}\left(\gamma\right)$
is $\sigma_{k,a}^{R}$ which can be used to capture the effect of
exogeneous stochastic processes like market dynamics on $R_{k,a}\left(\gamma\right)$.
The relative change between $R_{k,a}\left(\gamma\right)$ and $X_{k,a}\left(\gamma\right)$
is captured with correlation coefficent $\rho_{k}\in\left[0,1\right)$.
It captures how much a deviation of $X_{k,a}\left(\gamma\right)$
around its mean $\mu_{k,a}^{X}$ will effect the deviation of $R_{k,a}\left(\gamma\right)$
around its mean $h_{k}\left(\mu_{k,a}^{X}\right)$. A monotonic increasing
function, $h_{k}\left(\cdot\right)$, and a positive correlation coefficient,
$\rho_{k}$, are intuitive because from a statistical standpoint it
implies that an operator who serves more customer demand generates
higher revenue.

Let $R_{k,i}\left(\gamma\right)$ denote the revenue earned by the
$k^{th}$ operator if it is a Tier-$i$ operator in epoch $\gamma$.
Tier-1 serves customer demand using both licensed and opportunistic
access while Tier-2 operator serves its customer demand using opportunistic
access only. Hence,
\begin{eqnarray}
R_{k,1}\left(\gamma\right) & = & R_{k,lc}\left(\gamma\right)+R_{k,op}\left(\gamma\right)\label{eq:2.2.4.1}\\
R_{k,2}\left(\gamma\right) & = & R_{k,op}\left(\gamma\right)\label{eq:2.2.4.2}
\end{eqnarray}

Notice that since $R_{k,lc}\left(\gamma\right)$ and $R_{k,op}\left(\gamma\right)$
are Gaussian, $R_{k,1}\left(\gamma\right)$ and $R_{k,2}\left(\gamma\right)$
are Gaussian as well.\vspace{-0.5em}

\subsection{Spectrum Allocation Model\label{subsec:Allocation-Model}}

Licensed channels are allocated to the set of interested licensed
osperators, $\mathcal{S}_{L}$, through spectrum auctions. The auction
for epoch $\gamma$ happens at time slot $\left(\gamma-1\right)T+1$.
The set of interested licensed operators bids for licensed channels.
Let $V_{k}\left(\gamma\right)$ be the bid of the $k^{th}$ operator
in epoch $\gamma$. Intuitively, the bid of the $k^{th}$ operator
should depend on its valuation of a licensed channel. The value of
a licensed channel to the $k^{th}$ operator in epoch $\gamma$ is
$R_{k,lc}\left(\gamma\right)$, the revenue it can generate in an
epoch using the licensed channel. We capture this dependence between
$V_{k}\left(\gamma\right)$ and $R_{k,lc}\left(\gamma\right)$ using
a correlation coefficient $\omega_{k}\in\left[0,1\right)$ between
them. In our model, $V_{k}\left(\gamma\right)$ and $R_{k,lc}\left(\gamma\right)$
are jointly Gaussian. We have,\vspace{-0.5em}
\begin{equation}
\begin{bmatrix}V_{k}\left(\gamma\right)\\
R_{k,lc}\left(\gamma\right)
\end{bmatrix}\sim\mathcal{N}\left(\begin{bmatrix}\mu_{k,lc}^{R}\\
\mu_{k,lc}^{R}
\end{bmatrix},\begin{bmatrix}\left(\sigma_{k,lc}^{R}\right)^{2} & \omega{}_{k}\left(\sigma_{k,lc}^{R}\right)^{2}\\
\omega{}_{k}\left(\sigma_{k,lc}^{R}\right)^{2} & \left(\sigma_{k,lc}^{R}\right)^{2}
\end{bmatrix}\right)\label{eq:2.3.0}
\end{equation}

\noindent for all $\gamma$. In (\ref{eq:2.3.0}), $\mu_{k,lc}^{R}=h_{k}\left(\widetilde{\mu}_{k,lc}^{x}T\right)$
(refer to (\ref{eq:2.2.2}) and (\ref{eq:2.2.3})). Using a stochastic
model like (\ref{eq:2.3.0}) to capture the relation between $V_{k}\left(\gamma\right)$
and $R_{k,lc}\left(\gamma\right)$ leads to a more generic system
model because we can abstract away from the exact bid estimation strategies
of the operators which may rely on various market externalities.

\begin{figure}[t]
\begin{centering}
\includegraphics[scale=0.4]{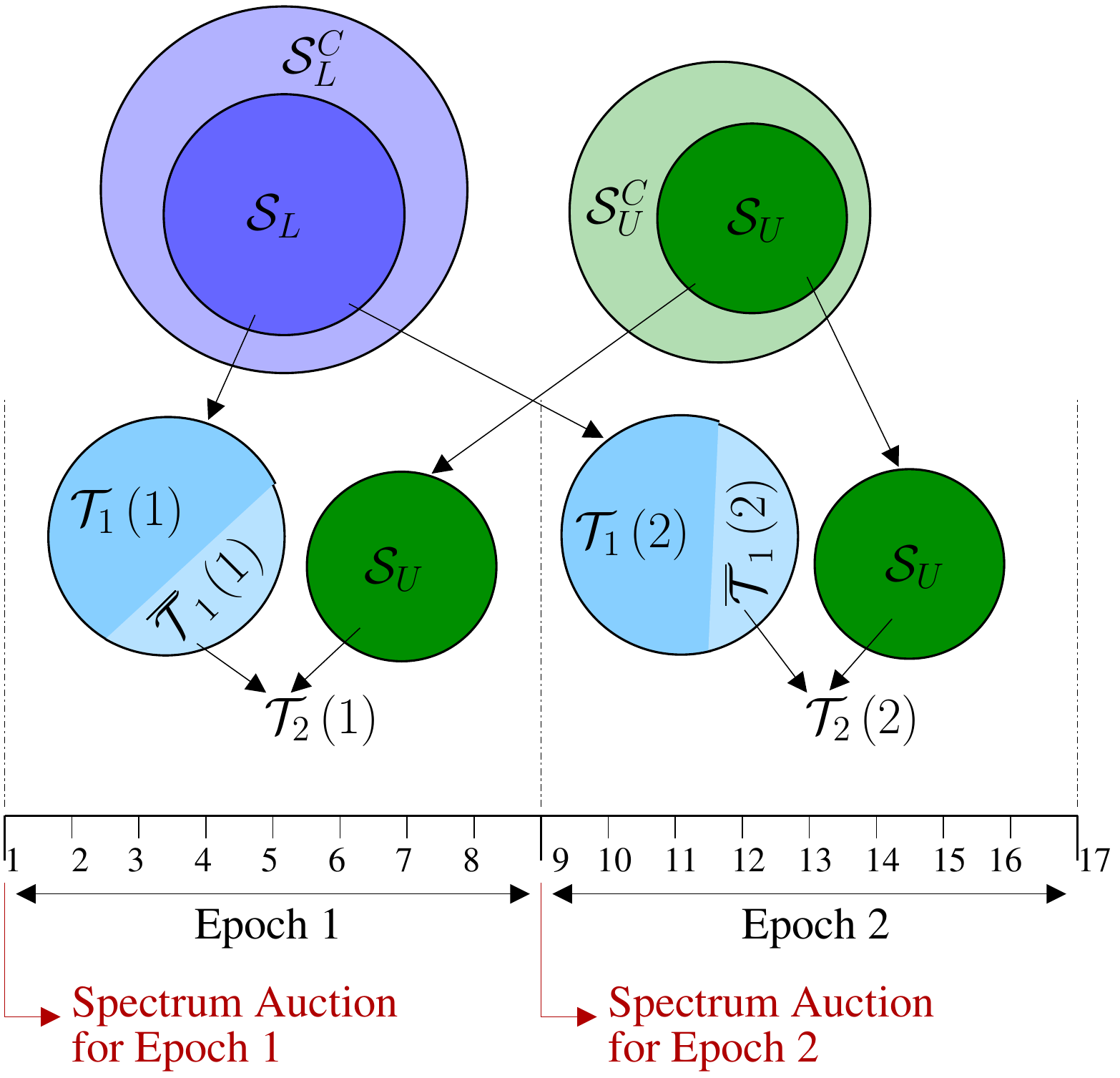}
\par\end{centering}
\caption{Pictorial representation of the set of candiate licensed operators
$\mathcal{S}_{L}^{C}$, candiate unlicensed operators $\mathcal{S}_{U}^{C}$,
interested licensed operators $\mathcal{S}_{L}$, interested unlicensed
operators $\mathcal{S}_{U}$, the set of interested licensed operators
who are allocated (not allocated) licensed channels in an epoch $\mathcal{T}_{1}\left(\gamma\right)$
($\overline{\mathcal{T}}_{1}\left(\gamma\right)$) and the set of
Tier-2 operators in an epoch $\mathcal{T}_{2}\left(\gamma\right)$.
Note that $\mathcal{T}_{1}\left(\gamma\right)$, $\overline{\mathcal{T}}_{1}\left(\gamma\right)$
and $\mathcal{T}_{2}\left(\gamma\right)$ are not same for epochs
$1$ and $2$.\vspace{-1.0em}\label{VariousSets}}
\end{figure}

Given that there are $P$ licensed channels and the spectrum cap is
one, the interested licensed operators with the $P$ highest bids
$V_{k}\left(\gamma\right)$ are allocated one licensed channel each
in epoch $\gamma$. The operators who are allocated a licensed channel
have to pay a price that is determined by the regulator. Such a pricing
model has been used in prior works \cite{constantleaseprice}. Let
$\mathcal{T}_{1}\left(\gamma\right)\subseteq\mathcal{S}_{L}$ denote
the set of interested licensed operators who are allocated licensed
channels in epoch $\gamma$. Similarly, $\overline{\mathcal{T}}_{1}\left(\gamma\right)=\mathcal{S}_{L}\backslash\mathcal{T}_{1}\left(\gamma\right)$
are the set of interested licensed operators who are not allocated
licensed channels in epoch $\gamma$. The operators in $\mathcal{T}_{1}\left(\gamma\right)$
serve their customer demand as Tier-1 operators in epoch $\gamma$.
On the other hand, operators in $\overline{\mathcal{T}}_{1}\left(\gamma\right)$
serve their customer demand as Tier-2 operators in epoch $\gamma$.
It is to be noted that $\mathcal{T}_{1}\left(\gamma\right)$ and $\overline{\mathcal{T}}_{1}\left(\gamma\right)$
are random sets as they get decided by the bids $V_{k}\left(\gamma\right)$
which are random variables. The set of Tier-2 operators in epoch $\gamma$
is $\mathcal{T}_{2}\left(\gamma\right)=\overline{\mathcal{T}_{1}}\left(\gamma\right)\bigcup\mathcal{S}_{U}$,
i.e., interested unlicensed operators and interested licensed operators
who are not allocated a licensed channel in epoch $\gamma$. Unlike
the sets $\mathcal{S}_{L}$ and $\mathcal{S}_{U}$ which are decided
once, sets $\mathcal{T}_{1}\left(\gamma\right)$, $\overline{\mathcal{T}}_{1}\left(\gamma\right)$,
and $\mathcal{T}_{2}\left(\gamma\right)$ are decided in the beginning
of every epoch. A pictorial representation of all the important sets
discussed till now is shown in Figure \ref{VariousSets}. Figure \ref{VariousSets}
also shows $\mathcal{T}_{1}\left(\gamma\right)$, $\overline{\mathcal{T}}_{1}\left(\gamma\right)$,
and $\mathcal{T}_{2}\left(\gamma\right)$ varies with epoch $\gamma$.\vspace{-1.0em}

\noindent 

Opportunistic spectrum allocation happens in every time slot to all
the Tier-2 operators. Tier-1 operators may or may not participate
in opportunistic spectrum access depending on whether $\phi$ is one
or zero. In order to capture both these cases under a single mathematical
abstraction, we modify the demand of Tier-1 and Tier-2 operators.
Let $\overline{x}_{k}\left(t\right)$ be the modified demand of the
$k^{th}$ operator which needs to be served using OSA. For time slot
$t$ of epoch $\gamma$,\vspace{-0.7em}

\begin{equation}
\overline{x}_{k}\left(t\right)=\begin{cases}
\phi\cdot\left(x_{k}\left(t\right)-\frac{D}{M}\right)^{+} & ;\,k\in\mathcal{T}_{1}\left(\gamma\right)\\
x_{k}\left(t\right) & ;\,k\in\mathcal{T}_{2}\left(\gamma\right)
\end{cases}\label{eq:2.3.1.0}
\end{equation}

According to (\ref{eq:2.3.1.0}), for a Tier-2 operator, its entire
demand $x_{k}\left(t\right)$ needs to be served using OSA. For Tier-1
operators, the excess demand which could not be satisfied with licensed
use is $\left(x_{k}\left(t\right)-\frac{D}{M}\right)^{+}$. If $\phi=1$,
this excess demand has to served using OSA. If $\phi=0$, then $\overline{x}_{k}\left(t\right)=0$
for Tier-1 operators implying that they don't participate in OSA.

Opportunistic channel capacity in time slot $t$ of epoch $\gamma$
is\vspace{-1.0em}

\begin{equation}
{\textstyle D_{O}\left(t\right)=\alpha_{U}\left(M-\widetilde{P}\right)\frac{D}{M}+\underset{k\in\mathcal{T}_{1}\left(\gamma\right)}{\sum}\mathcal{C}\left(x_{k}\left(t\right),\alpha_{L}\right)}\label{eq:2.3.1}
\end{equation}

\noindent where $\widetilde{P}=\min\left(\left|\mathcal{S}_{L}\right|,P\right)$.
In (\ref{eq:2.3.1}), the first term is the net channel capacity of
unlicensed channels and the second term is the net remaining channel
capacity of the licensed channels. The variable $\widetilde{P}$ is
used to capture edge cases where the number of licensed channels is
more than number of interested licensed operators. In such cases,
the remaining $P-\left|\mathcal{S}_{L}\right|$ channels which are
not allocated to licensed operators are used as unlicensed channels.
The expression for $\mathcal{C}\left(x_{k}\left(t\right),\alpha_{L}\right)$
depends on the OSA strategy (overlay or interweave) and has been discussed
in Section \ref{subsec:Channel-Model}. As our model is inspired by
the CBRS band, we have to ensure that opportunistic spectrum allocation
is fair \cite{CBRSFair}. One approach to ensure fair allocation and
to avoid wastage of channel capacity is to use a max-min fair algorithm,
like the famous Waterfilling algorithm. A detailed exposition of max-min
fairness can be found in \cite{MaxMinFair_Bertsekas,MaxMinFair_TON}.
In this section, we present the Waterfilling algorithm, explain its
working with an example and qualitatively justify how it ensures fairness
and avoids wastage of channel capacity. Waterfilling algorithm will
be used for opportunistic channel allocation throughout this paper.

\begin{figure}[t]
\begin{centering}
\includegraphics[scale=0.48]{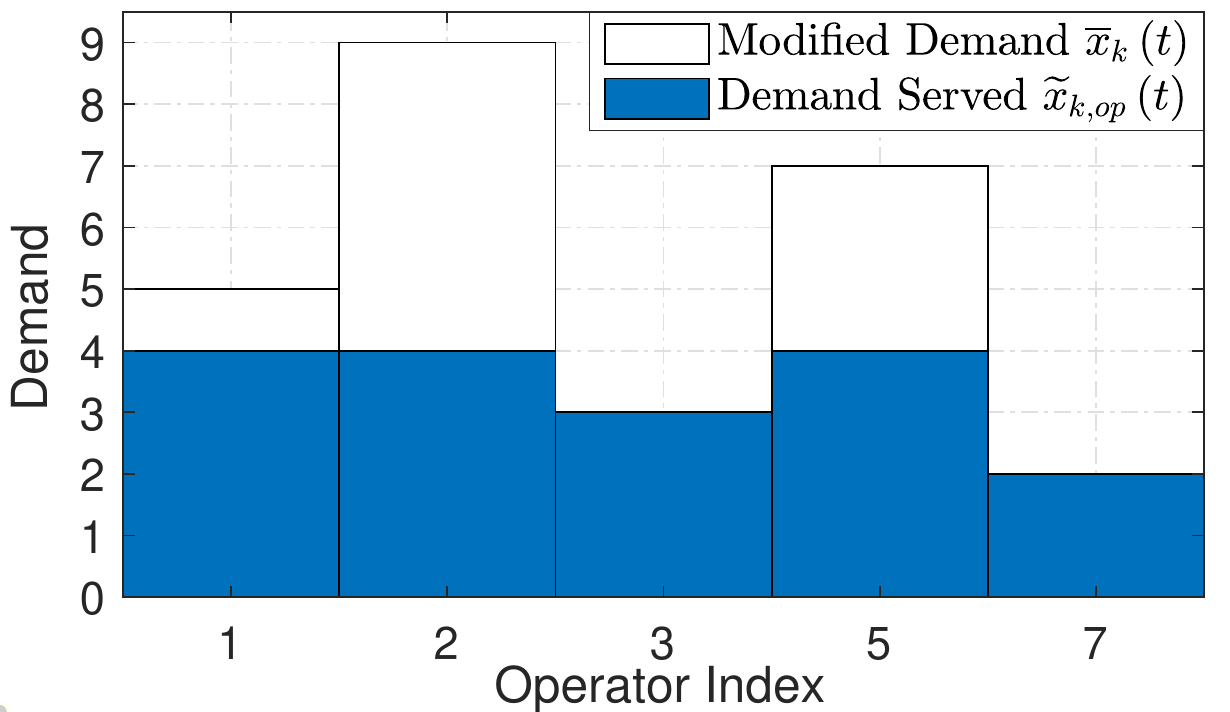}
\par\end{centering}
\caption{Pictorial representation of the example for Waterfilling algorithm.\vspace{-3.0em}}
\label{waterfilling_example}
\end{figure}

Algorithm \ref{WaterFilling_algo} is the psuedocode of Waterfilling
algorithm. Let $\mathcal{S}$ denote the set of interested operators,
i.e. $\mathcal{S}=\mathcal{S}_{L}\bigcup\mathcal{S}_{C}$. The union
of Tier-1 and Tier-2 operators, $\mathcal{T}_{1}\left(\gamma\right)\bigcup\mathcal{T}_{2}\left(\gamma\right)$,
is equal to $\mathcal{S}$. The inputs to Algorithm \ref{WaterFilling_algo}
are the opportunistic channel capacity, $D_{O}\left(t\right)$, and
the modified demands of all the interested operators, $\ensuremath{\left\{ \overline{x}_{k}\left(t\right)\right\} _{k\in\mathcal{S}}}$.
The output of Algorithm \ref{WaterFilling_algo} is the opportunistic
channel capacity allocated to all the interested operators, $\ensuremath{\left\{ \widetilde{x}_{k,op}\left(t\right)\right\} _{k\in\mathcal{S}}}$.
$\widetilde{x}_{k,op}\left(t\right)$ is also equal to the demand
served by the operators using OSA. We use the following example to
explain Algorithm \ref{WaterFilling_algo}: the set of interested
operators is $\mathcal{S}=\left\{ 1,2,3,5,7\right\} $, their corresponding
modified demand is $\left\{ 5,9,3,7,2\right\} $, and the opportunistic
channel capacity $D_{O}\left(t\right)=17$. The example is shown in
Figure \ref{waterfilling_example}.\vspace{-1.0em}

\noindent \SetInd{0.1em}{0.75em}
\SetAlgoHangIndent{0.0em}
\begin{algorithm}[t]

	\DontPrintSemicolon
	 
	\SetKwInput{Input}{Input}
	\SetKwInput{Output}{Output}

	\Input{$D_{O}\left(t\right)$, $\left\{ \overline{x}_{k}\left(t\right)\right\} _{k\in\mathcal{S}}$}

	\Output{$\ensuremath{\left\{ \widetilde{x}_{k,op}\left(t\right)\right\} _{k\in\mathcal{S}}}$}

	\justfy Sort the list $\left\{ \overline{x}_{k}\left(t\right)\right\} _{k\in\mathcal{S}}$ in ascending order of $\overline{x}_{k}\left(t\right)$. Let $\kappa\left(j\right)$  denote the operator index corresponding to the $j^{th}$ position of the sorted list.

	\justfy Set unused opportunistic channel capacity $C = D_{O}\left(t\right)$ and the remaining number of interested operators to allocate   channel capacity $N_{S}=\left|\mathcal{S}\right|$.

	\justfy \For{$j\leftarrow 1$ \KwTo $\left|\mathcal{S}\right|$}
	{
		\justfy Set $\widetilde{x}_{\kappa\left(j\right),op}\left(t\right)=\min\left(\overline{x}_{\kappa\left(j\right)}\left(t\right)\,,\,\frac{C}{N_{S}}\right)$.

		\justfy Set $C=C-\widetilde{x}_{\kappa\left(j\right),op}\left(t\right)$ and $N_{S}=N_{S}-1$.
	}
			
	\caption{Waterfilling Algorithm for Opportunistic Channel Allocation}
	\label{WaterFilling_algo}
\end{algorithm}

Waterfilling algorithm allocates channel capacity to the set of interested
operators in ascending order of their modified demand (\textit{lines
1} and \textit{3}). The sorted list of modified demand is $\left\{ 2,3,5,7,9\right\} $
and the operator index $\kappa\left(j\right)$ corresponding to position
$j=1,2,3,4,5$ of the sorted list is $7$, $3$, $1$, $5$, $2$
respectively. In \textit{line 2}, unused opportunistic channel capacity
$C=17$ and the remaining number of interested operators who needs
to be allocated channel capacity $N_{S}=5$. Inside the for loop,
the algorithm reserves equal portion of unused opportunistic channel
capacity $C$ for the remaining $N_{S}$ interested operators. This
is done in \textit{line 4} where a maximum channel capacity of $\frac{C}{N_{S}}$
is reserved for the $\kappa\left(j\right)^{th}$ operator. This step
ensures \textit{fairness} of Waterfilling algorithm. The channel capacity
allocated to the $\kappa\left(j\right)^{th}$ operator is the minimum
of its modified demand (the required channel capacity) and the maximum
reserved channel capacity of $\frac{C}{N_{S}}$. Accordingly, $C$
and $N_{S}$ are updated in \textit{line 5}. In our example, for $j=1$,
$\widetilde{x}_{7,op}\left(t\right)=\min\left(2,\frac{17}{5}\right)=2$
and hence the updated $C=17-2=15$ and $N_{S}=4$. For $j=2$, $\widetilde{x}_{3,op}\left(t\right)=\min\left(3,\frac{15}{4}\right)=3$
and hence the updated $C=15-3=12$ and $N_{S}=3$. For $j=3$, $\widetilde{x}_{1,op}\left(t\right)=\min\left(5,\frac{12}{3}\right)=4$
and hence the updated $C=12-4=8$ and $N_{S}=2$. The loop continues
and finally $\widetilde{x}_{5,op}\left(t\right)=\widetilde{x}_{2,op}\left(t\right)=4$.
Waterfilling algorithm \textit{prevents wastage} of channel capacity
by allocating no more than required channel capacity in \textit{line
4}. This ensures that the unused opportunistic channel capacity $C$
is as high as possible for the operators with higher customer demand.

We end this section by proving that the output $\widetilde{x}_{k,op}\left(t\right)$
of Algorithm \ref{WaterFilling_algo} are iid random variables. By
refering to (\ref{eq:2.3.1}) and (\ref{eq:2.3.1.0}), we can conclude
that $D_{O}\left(t\right)$ and $\overline{x}_{k}\left(t\right)$,
which forms the input to Algorithm \ref{WaterFilling_algo}, are the
ouptuts of time invariant functions of iid random variables $x_{k}\left(t\right)$
and $V_{k}\left(\gamma\right)$ ($V_{k}\left(\gamma\right)$ decides
the random set $\mathcal{T}_{1}\left(\gamma\right)$ in (\ref{eq:2.3.1})).
This implies that $D_{O}\left(t\right)$ and $\overline{x}_{k}\left(t\right)$
are iid random variables as well. Also note that except the inputs
$D_{O}\left(t\right)$ and $\overline{x}_{k}\left(t\right)$, Algorithm
\ref{WaterFilling_algo} is not dependent on time $t$. Therefore,
Algorithm \ref{WaterFilling_algo} can be expressed as a time-invariant
function of iid random variables $D_{O}\left(t\right)$ and $\overline{x}_{k}\left(t\right)$.
This directly implies that the output $\widetilde{x}_{k,op}\left(t\right)$
of Algorithm \ref{WaterFilling_algo} are iid random variables.\vspace{-1.0em}

\noindent 

\textit{Remark 2: Generality of the OSA model.} We want to highlight
that our OSA model is very general for three reasons. \textit{First},
any opportunistic channel allocation algorithm can be used as long
as $\widetilde{x}_{k,op}\left(t\right)$ are iid random variables.
\textit{Second}, the parameter $\phi$ helps us capture cases where
Tier-1 operators can/cannot participate in OSA. \textit{Third}, our
model can capture both overlay and interweave OSA strategy.

\section{Optimization Problem\label{sec:Optimization-Problem}}

We start this section by formulating the optimization problem for
joint spectrum partition as a two- stage Stackelberg Game in Section
\ref{subsec:Stackelberg}. In the process of formulating the Stackelberg
Game, we introduce two functions. \textit{First}, is the revenue function
of an operator which captures the expected revenue of an operator
in an epoch. \textit{Second}, is the objective function which is proportional
to spectrum utilization of all the interested operators in the market.
We then develop efficient algorithms to solve the two stages of the
Stackelberg Game in Section \ref{subsec:Solution} and hence find
the optimal $M$ and $P$ which maximizes spectrum utilization. In
sections \ref{subsec:Stackelberg} and \ref{subsec:Solution}, we
assume \textit{complete information} games. This assumption leads
to notational simplicity. Also, this assumption does not affect the
technical contribution of the paper as the overall approach can be
easily extended to \textit{incomplete information} games. This is
discussed in Section \ref{subsec:Incomplete-Information}.\vspace{-0.5em}

\subsection{Stackelberg Game Formulation\label{subsec:Stackelberg}}

In this section, we formulate the optimal spectrum partitioning problem
as a two-stage Stackelberg game between the regulator and the wireless
operators. The $k^{th}$ operator can be completely characterised
by seven parameters which can be represented as a tuple $\xi_{k}=\left(\mu_{k}^{\theta},\,\sigma_{k}^{\theta},\,h_{k}\left(\cdot\right),\,\sigma_{k,a}^{R},\,\rho_{k},\,\omega_{k},\,\lambda_{k}\right)$.
In sections \ref{subsec:Stackelberg} and \ref{subsec:Solution},
we assume complete information games, i.e. an operator and the regulator
knows $\xi_{k}$ of all the operators. The player in stage-1 of the
Stackelberg game is the regulator whose decision variables are $M$
and $P$. The payoff of the regulator is the expected spectrum utilization
over a period of $\Gamma\geq1$ epochs which is given by\vspace{-1.0em}

\begin{equation}
{\textstyle {\displaystyle Q=}E\left[\stackrel[\gamma=1]{\Gamma}{\sum}\;\stackrel[t=\left(\gamma-1\right)T+1]{\gamma T}{\sum}\left(Q_{lc}\left(\gamma,t\right)+Q_{op}\left(\gamma,t\right)\right)\right]}\;\text{{where,}}\label{eq:2.4.1}
\end{equation}
\vspace{-1.0em}
\begin{equation}
Q_{lc}\left(\gamma,t\right)=\underset{k\in\mathcal{T}_{1}\left(\gamma\right)}{{\textstyle \sum}}\widetilde{x}_{k,lc}\left(t\right)\label{eq:2.4.2}
\end{equation}
\vspace{-0.5em}
\begin{equation}
Q_{op}\left(\gamma,t\right)=\underset{k\in\mathcal{S}}{{\textstyle \sum}}\widetilde{x}_{k,op}\left(t\right)\label{eq:2.4.3}
\end{equation}

In (\ref{eq:2.4.1}), the inner summation is to calculate spectrum
utilization over all time slots in an epoch while the outer summation
is to calculate spectrum utilization over all the epochs. $Q_{lc}\left(\gamma,t\right)$
and $Q_{op}\left(\gamma,t\right)$ are the net spectrum utilization
in time slot $t$ of epoch $\gamma$ by using licensed and opportunistic
spectrum access respectively. The regulator wants to maximize $Q$.
Using \textit{linearity of expectation}, we can rewrite (\ref{eq:2.4.1})
as\vspace{-1.0em}

\begin{equation}
Q=\stackrel[\gamma=1]{\Gamma}{\sum}\;\stackrel[t=\left(\gamma-1\right)T+1]{\gamma T}{\sum}E\left[Q_{lc}\left(\gamma,t\right)+Q_{op}\left(\gamma,t\right)\right]\label{eq:2.4.4}
\end{equation}

We now prove that $E\left[Q_{lc}\left(\gamma,t\right)+Q_{op}\left(\gamma,t\right)\right]$
is not a function of $\gamma$ and $t$. Based on (\ref{eq:2.4.2})
and (\ref{eq:2.4.3}), $\widetilde{x}_{k,lc}\left(t\right)$, $\widetilde{x}_{k,op}\left(t\right)$,
and $\mathcal{T}_{1}\left(\gamma\right)$ are the only random variables
in the expressions of $Q_{lc}\left(\gamma,t\right)$ and $Q_{op}\left(\gamma,t\right)$.
As discussed in previous sections, $\widetilde{x}_{k,lc}\left(t\right)$
and $\widetilde{x}_{k,op}\left(t\right)$ are iid random variables.
$\mathcal{T}_{1}\left(\gamma\right)$ is a function of bids $V_{k}\left(\gamma\right)$
of the operators. Since, $V_{k}\left(\gamma\right)$ is an iid random
variable, so is $\mathcal{T}_{1}\left(\gamma\right)$. This discussion
implies that $Q_{lc}\left(\gamma,t\right)+Q_{op}\left(\gamma,t\right)$
itself is an iid random variable and hence its expectation is independent
of $\gamma$ and $t$. Infact, it is a function of $M$, $P$, $\mathcal{S}_{L}$
and $\mathcal{S}_{U}$. Let,\vspace{-1.0em}

\begin{equation}
U\left(M,P,\mathcal{S}_{L},\mathcal{S}_{U}\right)=E\left[Q_{lc}\left(\gamma,t\right)+Q_{op}\left(\gamma,t\right)\right]\label{eq:2.4.5}
\end{equation}

Substituting (\ref{eq:2.4.5}) in (\ref{eq:2.4.4}) we get,
\begin{equation}
Q=\Gamma TU\left(M,P,\mathcal{S}_{L},\mathcal{S}_{U}\right)\label{eq:2.4.6}
\end{equation}

Equation \ref{eq:2.4.6} shows that maximizing $Q$ is same as maximizing
$U\left(M,P,\mathcal{S}_{L},\mathcal{S}_{U}\right)$. Therefore, we
will use $U\left(M,P,\mathcal{S}_{L},\mathcal{S}_{U}\right)$ as the
payoff function of the regulator in the rest of the paper. $U\left(M,P,\mathcal{S}_{L},\mathcal{S}_{U}\right)$
is also called the \textit{objective function} as it is a direct measure
of spectrum utilization which we are trying to maximize in this paper.

The players in stage-2 of the Stackelberg game are the candidate licensed
operators, $\mathcal{S}_{L}^{C}$, and candidate unlicensed operators,
$\mathcal{S}_{U}^{C}$. The decision variables of the Stage-2 game
are the set of interested licensed operators, $\mathcal{S}_{L}$,
and the set of interested unlicensed operators, $\mathcal{S}_{U}$.
The $k^{th}$ operator is interested in joining the market only if
the expected revenue in an epoch is greater than $\lambda_{k}$. The
expected revenue in an epoch of an interested licensed or unlicensed
operator is given by the\textit{ revenue function}. The formula for
revenue function if different for interested licensed operators and
interested unlicensed operators. The revenue function of an interested
licensed operator, i.e. $k\in\mathcal{S}_{L}$, is\vspace{-1.0em}

\[
\mathcal{R}_{k}\left(M,P,\mathcal{S}_{L},\mathcal{S}_{U}\right)\quad\quad\quad\quad\quad\quad\quad\quad\quad\quad\quad\quad\quad\quad
\]
\vspace{-2.5em}

\begin{equation}
=E\left[R_{k,1}\left(\gamma\right)\,|\,\mathcal{E}_{k}^{\gamma}\right]P\left[\mathcal{E}_{k}^{\gamma}\right]+E\left[R_{k,2}\left(\gamma\right)\,|\,\overline{\mathcal{E}_{k}^{\gamma}}\right]P\left[\overline{\mathcal{E}_{k}^{\gamma}}\right]\label{eq:2.4.7}
\end{equation}
\vspace{-2.0em}

\[
=E\left[R_{k,lc}\left(\gamma\right)\,|\,\mathcal{E}_{k}^{\gamma}\right]P\left[\mathcal{E}_{k}^{\gamma}\right]\quad\quad\quad\quad\quad\quad\quad\quad\quad\quad\quad\quad\quad
\]
\vspace{-2.5em}

\begin{equation}
+E\left[R_{k,op}\left(\gamma\right)\,|\,\mathcal{E}_{k}^{\gamma}\right]P\left[\mathcal{E}_{k}^{\gamma}\right]+E\left[R_{k,op}\left(\gamma\right)\,|\,\overline{\mathcal{E}_{k}^{\gamma}}\right]P\left[\overline{\mathcal{E}_{k}^{\gamma}}\right]\label{eq:2.4.7.1}
\end{equation}
\vspace{-2.0em}

\begin{equation}
=E\left[R_{k,lc}\left(\gamma\right)\,|\,\mathcal{E}_{k}^{\gamma}\right]P\left[\mathcal{E}_{k}^{\gamma}\right]+h_{k}\left(\mu_{k,op}^{X}\right)\quad\quad\quad\quad\quad\quad\label{eq:2.4.7.2}
\end{equation}

\noindent where $P\left[Z\right]$ denotes the probability of event
$Z$, $\mathcal{E}_{k}^{\gamma}$ ($\overline{\mathcal{E}_{k}^{\gamma}}$)
is the event that $k\in\mathcal{T}_{1}\left(\gamma\right)$ ($k\in\mathcal{T}_{2}\left(\gamma\right)$).
In (\ref{eq:2.4.7}), $E\left[R_{k,i}\left(\gamma\right)\,|\,\mathcal{E}_{k}^{\gamma}\right]$
is the expected revenue of the $k^{th}$ operator if it is a Tier-$i$
operator in epoch $\gamma$. Finally, (\ref{eq:2.4.7}) is obtained
using the \textit{law of total expectation}. Equation \ref{eq:2.4.7.1}
is obtained by substituting $R_{k,1}\left(\gamma\right)=R_{k,lc}\left(\gamma\right)+R_{k,op}\left(\gamma\right)$
(refer to (\ref{eq:2.2.4.1})). Equation \ref{eq:2.4.7.2} is obtained
by noticing that the sum of the second and the third term of (\ref{eq:2.4.7.1})
is equal to $E\left[R_{k,op}\left(\gamma\right)\right]$ which in
turn in equal to $h_{k}\left(\mu_{k,op}^{X}\right)$ according to
(\ref{eq:2.2.3}). Similar to the objective function, the revenue
function of an interested licensed operator is also not a function
of epoch $\gamma$. This is because the statistical properties of
the involved random variables $R_{k,1}\left(\gamma\right)$ and $R_{k,2}\left(\gamma\right)$
are independent of $\gamma$.

If the $k^{th}$ operator is an interested unlicensed operator, i.e.
$k\in\mathcal{S}_{U}$, it is always a Tier-2 operator. Hence, its
expected revenue in an epoch is
\begin{equation}
\mathcal{R}_{k}\left(M,P,\mathcal{S}_{L},\mathcal{S}_{U}\right)=E\left[R_{k,2}\left(\gamma\right)\right]=h_{k}\left(\mu_{k,op}^{X}\right)\label{eq:2.4.8}
\end{equation}
where $E\left[R_{k,2}\left(\gamma\right)\right]=h_{k}\left(\mu_{k,op}^{X}\right)$
because $R_{k,2}\left(\gamma\right)=R_{k,op}\left(\gamma\right)$
(refer to (\ref{eq:2.2.4.2})) and $E\left[R_{k,op}\left(\gamma\right)\right]=h_{k}\left(\mu_{k,op}^{X}\right)$.

Payoff function of an operator who is interested in joining the market
either as a licensed or an unlicensed operator is\vspace{-1.2em}

\begin{equation}
\pi_{k}\left(M,P,\mathcal{S}_{L},\mathcal{S}_{U}\right)=\mathcal{R}_{k}\left(M,P,\mathcal{S}_{L},\mathcal{S}_{U}\right)-\lambda_{k}\label{eq:2.4.9}
\end{equation}

\noindent where $\mathcal{R}_{k}\left(M,P,\mathcal{S}_{L},\mathcal{S}_{U}\right)$
is given by (\ref{eq:2.4.7}) if $k\in\mathcal{S}_{L}$ and by (\ref{eq:2.4.8})
if $k\in\mathcal{S}_{U}$. If an operator does not join the market,
its payoff is zero. An operator decides to enter the market only if
its payoff $\pi_{k}\left(M,P,\mathcal{S}_{L},\mathcal{S}_{U}\right)$
is \textit{strictly} greater than zero.

With (\ref{eq:2.4.9}) as the payoff function, Stage-2 game may have
multiple Nash Equilibriums which complicates the analysis. This can
be simplified if we assume that the operators are \textit{pessimistic}
in nature. By doing so, we can get an unique solution of the Stage-2
game. Pessimistic models to address the issue of multiple Nash Equilibriums
have been considered in prior works like \cite{pessimistic1,pessimistic2,pessimistic3}.
One simple approach to model pessimistic decision making strategy
is to use the concept of \textit{dominant strategy}, i.e. an operator
decides to join the market if and only if joining the market is its
optimal strategy irrespective of whether other operators decides to
join the market. However, in this paper, we model a pessimistic operators'
decision making strategy using \textit{iterated elimination of strictly
dominated strategies} (IESDS) \cite{game_theory_shoham}. Compared
to dominant strategy, IESDS is a less pessimistic decision making
strategy because more operators will join the market.

IESDS can be explained as follows. IESDS consists of iterations. Consider
the first iteration which is the original Stage-2 game. We iterate
through all the candidate licensed and unlicensed operators to check
if either joining the market or not joining the market is a dominant
strategy for any of the operators. The operators whose dominant strategy
is to join (not join) the market will join (not join) the market irrespective
of other operators' decisions. This reduces the size of the Stage-2
game as it effectively consist of those operators who could not decide
whether to join (not join) the market in the first iteration. Such
operators are called\textit{ confused operators} in this paper. These
confused operators who did not have a dominant strategy in the original
Stage-2 game may have a dominant strategy in the reduced Stage-2 game.
Therefore, in the second iteration, we find the dominant strategy
of the confused operators in the reduced Stage-2 game. Such iterations
continue until convergence, which happens when the reduced Stage-2
game does not have any dominant strategy. It is possible that there
are 'confused operators' even after convergence. Those operators will
not join the market because in our model, the operators are pessimistic
in nature.\vspace{-0.5em}

\subsection{Solution of the Stackelberg Game\label{subsec:Solution}}

In this subsection, we use the process of backward induction \cite{mas1995microeconomic}
to solve the Stackelberg Game formulated in Section \ref{subsec:Stackelberg}.
To apply backward induction, we first solve Stage-2 of the game followed
by Stage-1. The following properties of the revenue function (as given
by (\ref{eq:2.4.7}) and (\ref{eq:2.4.8})) are crucial in designing
an efficient algorithm to solve Stage-2 of the Stackelberg Game.

\begin{property} \label{RevenueFuncProp1}
$\mathcal{R}_{k}\left(M,P,\mathcal{S}_{L},\mathcal{S}_{U}\right)$ is monotonic decreasing in $\mathcal{S}_{L}$, i.e. $\mathcal{R}_{k}\left(M,P,\mathcal{S}_{L},\mathcal{S}_{U}\right)\geq\mathcal{R}_{k}\left(M,P,\mathcal{S}_{L}\bigcup\left\{ a\right\} ,\mathcal{S}_{U}\right)$ where $a\notin\mathcal{S}_{L}$ and $a\in\mathcal{S}_{L}^{C}$.
\end{property}

\begin{property} \label{RevenueFuncProp2}
$\mathcal{R}_{k}\left(M,P,\mathcal{S}_{L},\mathcal{S}_{U}\right)$ is monotonic decreasing in $\mathcal{S}_{U}$, i.e. $\mathcal{R}_{k}\left(M,P,\mathcal{S}_{L},\mathcal{S}_{U}\right)\geq\mathcal{R}_{k}\left(M,P,\mathcal{S}_{L},\mathcal{S}_{U}\bigcup\left\{ a\right\} \right)$ where $a\notin\mathcal{S}_{U}$ and $a\in\mathcal{S}_{U}^{C}$.
\end{property}

We have verified these properties numerically using the Monte Carlo
integrator which will described in Section \ref{sec:Monte-Carlo-Integrator}.
These properties can be justified as follows. Property \ref{RevenueFuncProp1}
states that as the set of interested licensed operators, $\mathcal{S}_{L}$,
increases, the revenue function of both the licensed and the unlicensed
operators decreases. The revenue function of a licensed operator decreases
with increase in $\mathcal{S}_{L}$ because the operator has to compete
with more operators in the spectrum auctions to get a channel. This
reduces the operator's probability of winning spectrum auctions which
in turn decreases its revenue function as it can effectively serve
fewer customer demand. The revenue function of an unlicensed operator
also decreases with increase in $\mathcal{S}_{L}$. This happens because
with increase in $\mathcal{S}_{L}$, there is an increase in the number
of operators interested in opportunistic channel access. This reduces
the share of opportunistic channels for an unlicensed operator. Therefore,
its revenue decreases as it can serve fewer customer demand. Property
\ref{RevenueFuncProp2} states that as the set of interested unlicensed
operators, $\mathcal{S}_{U}$, increases, the revenue function of
both the licensed and the unlicensed operators decreases. This happens
because with increase in $\mathcal{S}_{U}$, the share of opportunistic
channel decreases for a licensed or an unlicensed operator. This in
turn decreases its revenue function.

\noindent \SetInd{0.0em}{0.55em}
\SetAlgoHangIndent{0.0em}
\begin{algorithm}[t]

	\DontPrintSemicolon
	 
	\SetKwInput{Input}{Input}
	\SetKwInput{Output}{Output}

	\Input{$M$, $P$, $T$, $D$, $\alpha_{L}$, $\alpha_{U}$, $\mathcal{S}_{L}^{C}$, $\mathcal{S}_{U}^{C}$ and $\xi_{k};\forall k\in\mathcal{S}_{L}^{C}\bigcup\mathcal{S}_{U}^{C}$}

	\Output{$\mathcal{S}_{L}\left(M,P\right)$ and $\mathcal{S}_{U}\left(M,P\right)$}

	\justfy Set $\widehat{\mathcal{X}}_{0}=\emptyset$, $\widetilde{\mathcal{X}}_{0}=\mathcal{S}_{L}^{C}$, $\widehat{\mathcal{Y}}_{0}=\emptyset$ and $\widetilde{\mathcal{Y}}_{0}=\mathcal{S}_{U}^{C}$.\;

	\justfy Set $converged = False$ and $l = 0$.\;

	\justfy \While{$not\left(converged\right)$}
	{
		\justfy Set $converged = True$ and $l=l+1$.\;

		\justfy Set $\widehat{\mathcal{X}}_{l}=\widehat{\mathcal{X}}_{l-1}$, $\widetilde{\mathcal{X}}_{l}=\widetilde{\mathcal{X}}_{l-1}$, $\widehat{\mathcal{Y}}_{l}=\widehat{\mathcal{Y}}_{l-1}$ and $\widetilde{\mathcal{Y}}_{l}=\widetilde{\mathcal{Y}}_{l-1}$.\;

		\justfy \For{$k$ \textbf{in} $\widetilde{\mathcal{X}}_{l-1}$}
		{
			\justfy \uIf{$\mathcal{R}_{k}\left(M,P,\widehat{\mathcal{X}}_{l-1}\bigcup\widetilde{\mathcal{X}}_{l-1},\widehat{\mathcal{Y}}_{l-1}\bigcup\widetilde{\mathcal{Y}}_{l-1}\right)>\lambda_{k}$}
			{
				\justfy Set $\widehat{\mathcal{X}}_{l}=\widehat{\mathcal{X}}_{l}\bigcup\left\{k\right\}$ and $\widetilde{\mathcal{X}}_{l}=\widetilde{\mathcal{X}}_{l}\backslash\left\{k\right\}$.\;

				\justfy Set $converged = False$.\;
			}

			\justfy \uElseIf{$\mathcal{R}_{k}\left(M,P,\widehat{\mathcal{X}}_{l-1}\bigcup\left\{k\right\},\widehat{\mathcal{Y}}_{l-1}\right)\leq \lambda_{k}$}
			{
				\justfy Set $\widetilde{\mathcal{X}}_{l}=\widetilde{\mathcal{X}}_{l}\backslash\left\{k\right\}$.\;

				\justfy Set $converged = False$.\;
			}
		}

		\justfy \For{$k$ \textbf{in} $\widetilde{\mathcal{Y}}_{l-1}$}
		{
			\justfy \uIf{$\mathcal{R}_{k}\left(M,P,\widehat{\mathcal{X}}_{l-1}\bigcup\widetilde{\mathcal{X}}_{l-1},\widehat{\mathcal{Y}}_{l-1}\bigcup\widetilde{\mathcal{Y}}_{l-1}\right)>\lambda_{k}$}
			{
				\justfy Set $\widehat{\mathcal{Y}}_{l}=\widehat{\mathcal{Y}}_{l}\bigcup\left\{k\right\}$ and $\widetilde{\mathcal{Y}}_{l}=\widetilde{\mathcal{Y}}_{l}\backslash\left\{k\right\}$.\;

				\justfy Set $converged = False$.\;
			}

			\justfy \uElseIf{$\mathcal{R}_{k}\left(M,P,\widehat{\mathcal{X}}_{l-1},\widehat{\mathcal{Y}}_{l-1}\bigcup\left\{k\right\}\right)\leq \lambda_{k}$}
			{
				\justfy Set $\widetilde{\mathcal{Y}}_{l}=\widetilde{\mathcal{Y}}_{l}\backslash\left\{k\right\}$.\;

				\justfy Set $converged = False$.\;
			}
		}
	}

	\justfy Set $\mathcal{S}_{L}\left(M,P\right)=\widehat{\mathcal{X}}_{l}$ and $\mathcal{S}_{U}\left(M,P\right)=\widehat{\mathcal{Y}}_{l}$\;

	\caption{Algorithm to solve Stage 2 of the Stackelberg Game for Joint Spectrum Partitioning}
	\label{opalgo_spectrum_partitioning_stage2}
\end{algorithm}

\vspace{-1.0em}

The psuedocode to solve Stage-2 of the Stackelberg game is given in
Algorithm \ref{opalgo_spectrum_partitioning_stage2}. The inputs of
Algorithm \ref{opalgo_spectrum_partitioning_stage2} are clearly described
in Table \ref{notations}. Let $\mathcal{S}_{L}\left(M,P\right)$
and $\mathcal{S}_{U}\left(M,P\right)$ denote the set of interested
licensed and unlicensed operators if the entire bandwidth is divided
into $M$ channels out of which $P$ are licensed channels. $\mathcal{S}_{L}\left(M,P\right)$
and $\mathcal{S}_{U}\left(M,P\right)$ are the outputs of Algorithm
\ref{opalgo_spectrum_partitioning_stage2}. As mentioned in Section
\ref{subsec:Stackelberg}, $\mathcal{S}_{L}\left(M,P\right)$ and
$\mathcal{S}_{U}\left(M,P\right)$ are decided by the operators based
on IESDS. Algorithm \ref{opalgo_spectrum_partitioning_stage2} uses
Properties \ref{RevenueFuncProp1} and \ref{RevenueFuncProp2} to
compute $\mathcal{S}_{L}\left(M,P\right)$ and $\mathcal{S}_{U}\left(M,P\right)$
in polynomial time when an operator's decision making strategy to
join/not join the market is based on IESDS.

Let $\widehat{\mathcal{X}}_{l}$ and $\widetilde{\mathcal{X}}_{l}$,
where $\widehat{\mathcal{X}}_{l}\,,\,\widetilde{\mathcal{X}}_{l}\in\mathcal{S}_{L}^{C}$,
denote the set of licensed operators who are sure to join the market
and the set of confused licensed operators respectively till the $l^{th}$
iteration. Note that $\widehat{\mathcal{X}}_{l}$ and $\widetilde{\mathcal{X}}_{l}$
are disjoint sets and the set $\mathcal{S}_{L}^{C}\backslash\left(\widehat{\mathcal{X}}_{l}\bigcup\widetilde{\mathcal{X}}_{l}\right)$
consists of those licensed operators who are sure not to join the
market till the $l^{th}$ iteration. Similarly, $\widehat{\mathcal{Y}}_{l}$
and $\widetilde{\mathcal{Y}}_{l}$, where $\widehat{\mathcal{Y}}_{l}\,,\,\widetilde{\mathcal{Y}}_{l}\in\mathcal{S}_{U}^{C}$,
denote the set of unlicensed operators who decided to join the market
and the set of confused unlicensed operators till the $l^{th}$ iteration
respectively.

We will now explain the working of Algorithm \ref{opalgo_spectrum_partitioning_stage2}.
Algorithm \ref{opalgo_spectrum_partitioning_stage2} starts with iteration
$0$. Initially, none of the operators are sure whether to join the
market or not; all of them are confused. Hence, in iteration $0$,
we initialize $\widehat{\mathcal{X}}_{0}=\emptyset$ , $\widetilde{\mathcal{X}}_{0}=\mathcal{S}_{L}^{C}$,
$\widehat{\mathcal{Y}}_{l}=\emptyset$ and $\widetilde{\mathcal{Y}}_{0}=\mathcal{S}_{U}^{C}$
(\textit{line 1}). The \textit{while loop} in \textit{lines 3-19}
finds $\widehat{\mathcal{X}}_{l}$, $\widetilde{\mathcal{X}}_{l}$,
$\widehat{\mathcal{Y}}_{l}$ and $\widetilde{\mathcal{Y}}_{l}$ for
the $l^{th}$ iteration given $\widehat{\mathcal{X}}_{l-1}$, $\widetilde{\mathcal{X}}_{l-1}$,
$\widehat{\mathcal{Y}}_{l-1}$ and $\widetilde{\mathcal{Y}}_{l-1}$
of the $\left(l-1\right)^{th}$ iteration. Since the operators in
sets $\widehat{\mathcal{X}}_{l-1}$ and $\widehat{\mathcal{Y}}_{l-1}$
will surely join the market, we initialize $\widehat{\mathcal{X}}_{l}$
and $\widehat{\mathcal{Y}}_{l}$ to $\widehat{\mathcal{X}}_{l-1}$
and $\widehat{\mathcal{Y}}_{l-1}$ respectively in the beginning of
the $l^{th}$ iteration (\textit{line 2}). The set of confused licensed
and unlicensed operators , $\widetilde{\mathcal{X}}_{l}$ and $\widetilde{\mathcal{Y}}_{l}$,
are initialized to $\widetilde{\mathcal{X}}_{l-1}$ and $\widetilde{\mathcal{Y}}_{l-1}$
respectively in the beginning of the $l^{th}$ iteration (\textit{line
2}). In the \textit{for loop} in \textit{lines 6 - 12}, we check if
any licensed operator in set $\widetilde{\mathcal{X}}_{l-1}$ is sure
to either join or not join the market. Similarly, in the \textit{for
loop} in \textit{lines 13 - 19}, we check if any unlicensed operator
in set $\widetilde{\mathcal{Y}}_{l-1}$ is sure to either join or
not join the market.

We will now explain the working of the for loop in lines 6-12. The
largest possible set of interested licensed operators in the $\widetilde{l}^{th}$
iteration, for $\widetilde{l}\geq l$ is $\widehat{\mathcal{X}}_{l-1}\bigcup\widetilde{\mathcal{X}}_{l-1}$.
This is because the operators in set $\mathcal{S}_{L}^{C}\backslash\left(\widehat{\mathcal{X}}_{l-1}\bigcup\widetilde{\mathcal{X}}_{l-1}\right)$
are sure not to join the market till the $\left(l-1\right)^{th}$
iteration. Similarly, the largest possible set of interested unlicensed
operators in the $\widetilde{l}^{th}$ iteration, for $\widetilde{l}\geq l$
is $\widehat{\mathcal{Y}}_{l-1}\bigcup\widetilde{\mathcal{Y}}_{l-1}$.
Therefore, according to Properties \ref{RevenueFuncProp1} and \ref{RevenueFuncProp2},
the minimum revenue of the $k^{th}$ operator, where $k\in\widetilde{\mathcal{X}}_{l-1}$,
in the $\widetilde{l}^{th}$ iteration, for $\widetilde{l}\geq l$
is $\ensuremath{\mathcal{R}_{k}\left(M,P,\widehat{\mathcal{X}}_{l-1}\bigcup\widetilde{\mathcal{X}}_{l-1},\widehat{\mathcal{Y}}_{l-1}\bigcup\widetilde{\mathcal{Y}}_{l-1}\right)}$.
So if\vspace{-0.5em}

\[
\mathcal{R}_{k}\left(M,P,\widehat{\mathcal{X}}_{l-1}\bigcup\widetilde{\mathcal{X}}_{l-1},\widehat{\mathcal{Y}}_{l-1}\bigcup\widetilde{\mathcal{Y}}_{l-1}\right)>\lambda_{k}
\]

\noindent then joining the market becomes the dominant strategy of
the $k^{th}$ operator in the $l^{th}$ iteration. Therefore, in \textit{line
8}, we remove the $k^{th}$ operator from the set of confused licensed
operators and add it to the set of licensed operators who are sure
to join the market. If the $k^{th}$ operator, where $k\in\widetilde{\mathcal{X}}_{l-1}$,
joins the market, then the smallest possible set of interested licensed
and unlicensed operators in the $\widetilde{l}^{th}$ iteration, for
$\widetilde{l}\geq l$ are $\widehat{\mathcal{X}}_{l-1}\bigcup\left\{ k\right\} $
and $\widehat{\mathcal{Y}}_{l-1}$ respectively. Therefore, according
to Properties \ref{RevenueFuncProp1} and \ref{RevenueFuncProp2},
the maximum revenue of the $k^{th}$ operator, where $k\in\widetilde{\mathcal{X}}_{l-1}$,
in the $\widetilde{l}^{th}$ iteration, for $\widetilde{l}\geq l$
is $\ensuremath{\mathcal{R}_{k}\left(M,P,\widehat{\mathcal{X}}_{l-1}\bigcup\left\{ k\right\} ,\widehat{\mathcal{Y}}_{l-1}\right)}$.
So if\vspace{-0.5em}

\[
\mathcal{R}_{k}\left(M,P,\widehat{\mathcal{X}}_{l-1}\bigcup\left\{ k\right\} ,\widehat{\mathcal{Y}}_{l-1}\right)\leq\lambda_{k}
\]

\noindent then not joining the market becomes the dominant strategy
of the $k^{th}$ operator in the $l^{th}$ iteration. Therefore, in
\textit{line 11}, we remove the $k^{th}$ operator from the set of
confused licensed operators but we do not add it to the set of licensed
operators who are sure to join the market. The for loop in lines 13-19
work in a similar way to decide if any unlicensed operator in set
$\widetilde{\mathcal{Y}}_{l-1}$ is sure to either join or not join
the market.

The variable $converged$ which is declared in \textit{line 2} and
updated in \textit{lines 9, 12, 16, 19} decides when the while loop
terminates. This can be explained as follows. Say that few of the
confused operators in the $l^{th}$ iteration decides to not join
the market, i.e. if statements in \textit{lines 10} or \textit{17}
are $true$. In this case, $converged$ is set to $false$ and hence
the while loop continues. Since few of the operators decides not to
join the market in the $l^{th}$ iteration, then due to Properties
\ref{RevenueFuncProp1} and \ref{RevenueFuncProp2}, the revenue function
of the remaining confused operators in the $\left(l+1\right)^{th}$
iteration is more compared to their corresponding values in the $l^{th}$
iteration. Therefore, it is possible that for some of these confused
operators, joining the market becomes the dominant strategy in the
$\left(l+1\right)^{th}$ iteration. The opposite happens when few
of the confused operators in the $l^{th}$ iteration decides to join
the market. This discussion captures the fundamental idea behind IESDS.

Say that after the end of the $l_{o}^{th}$ iteration, $\widehat{\mathcal{X}}_{l_{o}}=\widehat{\mathcal{X}}_{l_{o}-1}$,
$\widetilde{\mathcal{X}}_{l_{o}}=\widetilde{\mathcal{X}}_{l_{o}-1}$,
$\widehat{\mathcal{Y}}_{l_{o}}=\widehat{\mathcal{Y}}_{l_{o}-1}$ and
$\widetilde{\mathcal{Y}}_{l_{o}}=\widetilde{\mathcal{Y}}_{l_{o}-1}$.
This happens when \textit{if statements} in \textit{lines 7, 10, 14,
17} are all $false$. When this happens, $converged$ is $true$ after
the end of the $l_{o}^{th}$ iteration and hence the while loop terminates.
This is because if $\widehat{\mathcal{X}}_{l_{o}}=\widehat{\mathcal{X}}_{l_{o}-1}$,
$\widetilde{\mathcal{X}}_{l_{o}}=\widetilde{\mathcal{X}}_{l_{o}-1}$,
$\widehat{\mathcal{Y}}_{l_{o}}=\widehat{\mathcal{Y}}_{l_{o}-1}$ and
$\widetilde{\mathcal{Y}}_{l_{o}}=\widetilde{\mathcal{Y}}_{l_{o}-1}$,
then the value of the revenue function in lines 7, 10, 14, 17 in the
$\left(l_{o}+1\right)^{th}$ iteration is same as that in $l_{o}^{th}$
iteration. Therefore, the if statements in lines 7, 10, 14 and 17
will be $false$ in the $\left(l_{o}+1\right)^{th}$ iteration just
like the $l_{o}^{th}$ iteration. This argument suggests that $\widehat{\mathcal{X}}_{l}=\widehat{\mathcal{X}}_{l_{o}}$,
$\widetilde{\mathcal{X}}_{l}=\widetilde{\mathcal{X}}_{l_{o}}$, $\widehat{\mathcal{Y}}_{l}=\widehat{\mathcal{Y}}_{l_{o}}$
and $\widetilde{\mathcal{Y}}_{l}=\widetilde{\mathcal{Y}}_{l_{o}}$
for all $l\geq l_{o}$ and hence convergence in $\widehat{\mathcal{X}}_{l}$
, $\widetilde{\mathcal{X}}_{l}$, $\widehat{\mathcal{Y}}_{l}$ and
$\widetilde{\mathcal{Y}}_{l}$ has been achieved. After convergence
is achieved, there are three kinds of operators. \textit{First}, the
operators in sets $\widehat{\mathcal{X}}_{l_{o}}$ and $\widehat{\mathcal{Y}}_{l_{o}}$
who are sure that they should join the market. \textit{Second}, the
operators in sets $\widehat{\mathcal{X}}_{l_{o}}\backslash\widetilde{\mathcal{X}}_{l_{o}}$
and $\widehat{\mathcal{Y}}_{l_{o}}\backslash\widetilde{\mathcal{Y}}_{l_{o}}$
who are sure that they should not join the market. \textit{Third},
the 'confused' operators in sets $\widetilde{\mathcal{X}}_{l_{o}}$
and $\widetilde{\mathcal{Y}}_{l_{o}}$. Since our model assumes that
the operators are pessimistic, confused operators will not join the
market. Hence, the set of interested licensed and unlicensed operators
are $\widetilde{\mathcal{X}}_{l_{o}}$ and $\widehat{\mathcal{Y}}_{l_{o}}$
respectively where $l_{o}$ is the last iteration of Algorithm \ref{opalgo_spectrum_partitioning_stage1}
(\textit{line 20}).
\begin{prop}
Time complexity of Algorithm \ref{opalgo_spectrum_partitioning_stage2}
is $\mathcal{O}\left(N^{2}\right)$ where $N=\left|\mathcal{S}_{L}^{C}\right|+\left|\mathcal{S}_{U}^{C}\right|$.
\end{prop}
\begin{IEEEproof}
The while loop continues until none of the confused operators of an
iteration have a dominant strategy. Such a condition is possible at
most $N$ times because there are only $N$ candidate operators. Hence,
the while loop is executed at most $N$ times. For a given iteration
of the while loop, the inner for loop in lines 6-12 is executed at
most $\left|\mathcal{S}_{L}^{C}\right|$ times and that in lines 13-19
is executed at most $\left|\mathcal{S}_{U}^{C}\right|$ times. Therefore,
the inner for loops runs at most $\left|\mathcal{S}_{L}^{C}\right|+\left|\mathcal{S}_{U}^{C}\right|=N$
times. This shows that the time complexity of Algorithm \ref{opalgo_spectrum_partitioning_stage2}
is $\mathcal{O}\left(N^{2}\right)$. This completes the proof.
\end{IEEEproof}
\textit{Remark 3: Efficiency of Algorithm }\ref{opalgo_spectrum_partitioning_stage2}.
Algorithm \ref{opalgo_spectrum_partitioning_stage2} uses Properties
\ref{RevenueFuncProp1} and \ref{RevenueFuncProp2} to decide whether
a confused operator will join the market or not by computing its revenue
function for the largest/smallest set of interested operators. Without
these properties, we have to compute the revenue function for an exponential
number of set of interested operators to decide whether a confused
operator will join the market or not.

\textit{Remark 4: Comparison with dominant strategy}. Only the $1^{st}$
iteration of Algorithm \ref{opalgo_spectrum_partitioning_stage2}
is required to find the dominant strategies of the operators. It is
for this reason that the set of interested operators, $\mathcal{S}_{L}\left(M,P\right)$
and $\mathcal{S}_{U}\left(M,P\right)$, will always be larger if operators'
decision making strategy is based on IESDS rather than dominant strategy.

Given that $\mathcal{S}_{L}\left(M,P\right)$ and $\mathcal{S}_{U}\left(M,P\right)$
are the solutions of the Stage-2 game, the objective function in (\ref{eq:2.4.5})
can be re-written as
\begin{equation}
\widetilde{U}\left(M,P\right)=U\left(M,P,\mathcal{S}_{L}\left(M,P\right),\mathcal{S}_{U}\left(M,P\right)\right)\label{eq:3.2.0}
\end{equation}

In Stage-1, the regulator chooses $M$ and $P$ to maximize $\widetilde{U}\left(M,P\right)$.
Let the optimal solution be $M^{*}$ and $P^{*}$, the optimal value
of the objective function be $U^{*}$, where $\mbox{{\ensuremath{U^{*}}=\ensuremath{\widetilde{U}\left(M^{*},P^{*}\right)}}}$,
and the optimal set of interested licensed and unlicensed operators
be $\mathcal{S}_{L}^{*}$ and $\mathcal{S}_{U}^{*}$, where $\mathcal{S}_{L}^{*}=\mathcal{S}_{L}\left(M^{*},P^{*}\right)$
and $\mathcal{S}_{U}^{*}=\mathcal{S}_{U}\left(M^{*},P^{*}\right)$.
$M^{*}$ and $P^{*}$ are found by performing a grid-search. The grid
search is detailed in Algorithm \ref{opalgo_spectrum_partitioning_stage1}.
As shown in \textit{lines 2 and 3} of Algorithm \ref{opalgo_spectrum_partitioning_stage1},
the grid search is performed from $M=1$ to a certain $M_{max}$ and
from $P=0$ to $\min\left(\left|\mathcal{S}_{L}^{C}\right|,M\right)$.
Note that since spectrum cap is one, the number of licensed channels
should be lesser than the number of candidate licensed operators,
$\left|\mathcal{S}_{L}^{C}\right|$. Time complexity of Algorithm
\ref{opalgo_spectrum_partitioning_stage1} is $\mathcal{O}\left(M_{max}\left|\mathcal{S}_{L}^{C}\right|\right)$.\vspace{-0.5em}

\subsection{Incomplete Information Stackelberg Game\label{subsec:Incomplete-Information}}

In this section, we discuss the generalization to the case of incomplete
information games where the $j^{th}$ operator has a point estimate
$\widehat{\xi}_{k}^{j}$ of the tuple $\xi_{k}$ of the $k^{th}$
operator. Let $\mathcal{S}^{C}=\mathcal{S}_{L}^{C}\bigcup\mathcal{S}_{U}^{C}$.
We have $j\in\left\{ 0\right\} \bigcup\mathcal{S}^{C}$ and $k\in\mathcal{S}^{C}$,
where $j=0$ is the index of the regulator. Also, $\widehat{\xi}_{k}^{k}=\xi_{k}$
because the $k^{th}$ operator knows its own tuple $\xi_{k}$. We
now discuss the steps involved in finding the optimal value of $M$,
$P$, and the objective function in this incomplete information setting.

The regulator decides the optimal values of $M$ and $P$. While in
the complete information case, the regulator exactly knows $\xi_{k};\forall k\in\mathcal{S}^{C}$,
it now only has an estimate $\widehat{\xi}_{k}^{0};\forall k\in\mathcal{S}^{C}$
in the incomplete information case. Therefore, as far as deciding
the optimal value of $M$ and $P$ is concerned, the regulator simply
uses $\widehat{\xi}_{k}^{0};\forall k\in\mathcal{S}^{C}$ as inputs
to Algorithms \ref{opalgo_spectrum_partitioning_stage2} and \ref{opalgo_spectrum_partitioning_stage1}.
Let $M^{*}$ and $P^{*}$ denote the optimal values of $M$ and $P$.

\noindent \SetInd{0.1em}{1.0em}
\SetAlgoHangIndent{0.0em}
\begin{algorithm}[t]

	\DontPrintSemicolon
	 
	\SetKwInput{Input}{Input}
	\SetKwInput{Output}{Output}

	\Input{$T$, $D$, $\alpha_{L}$, $\alpha_{U}$, $\mathcal{S}_{L}^{C}$, $\mathcal{S}_{U}^{C}$, and  $\xi_{k};\forall k\in\mathcal{S}_{L}^{C}\bigcup\mathcal{S}_{U}^{C}$}

	\Output{$M^{*}$, $P^{*}$, $\mathcal{S}^{*}_{L}$, $\mathcal{S}^{*}_{U}$, and $U^{*}$}

	\justfy Set $U^{*}=-\infty$.\;

	\justfy \For{$M\leftarrow 1$ \KwTo $M_{max}$}
	{
		\justfy \For{$P\leftarrow 0$ \KwTo $\min\left(\left|\mathcal{S}_{L}^{C}\right|,M\right)$}
		{
			\justfy Call Algorithm \ref{opalgo_spectrum_partitioning_stage2} to get the set of interested licensed and unlicensed operators, $\mathcal{S}_{L}\left(M,P\right)$ and $\mathcal{S}_{U}\left(M,P\right)$ respectively, for current $M$ and $P$.

			\justfy Set $\widetilde{U}=U\left(M,P,\mathcal{S}_{L}\left(M,P\right),\mathcal{S}_{U}\left(M,P\right)\right)$.\;

			\justfy \If{$\widetilde{U}>U^{*}$}
			{
				\justfy Set $M^{*}=M$, $P^{*}=M$, $\mathcal{S}^{*}_{L}=\mathcal{S}_{L}\left(M,P\right)$, $\mathcal{S}^{*}_{U}=\mathcal{S}_{U}\left(M,P\right)$, and $U^{*}=\widetilde{U}$.\;
			}
		}
	}
	\caption{Algorithm to solve Stage 1 of the Stackelberg Game for Joint Spectrum Partitioning}
	\label{opalgo_spectrum_partitioning_stage1}
\end{algorithm}

\vspace{-1.0em}

$\mathcal{S}_{L}\left(M^{*},P^{*}\right)$ and $\mathcal{S}_{U}\left(M^{*},P^{*}\right)$
obtained using $\widehat{\xi}_{k}^{0};\forall k\in\mathcal{S}^{C}$
as an input to Algorithm \ref{opalgo_spectrum_partitioning_stage2}
may not be the true set of interested licensed and unlicensed operators
corresponding to $M^{*}$ and $P^{*}$. This is because the estimate
$\widehat{\xi}_{k}^{j}$ varies between the regulator and the operators.
Let $\mathcal{A}_{L}^{j}=\mathcal{S}_{L}\left(M^{*},P^{*},\left\{ \widehat{\xi}_{k}^{j}\right\} _{k\in\mathcal{S}^{C}}\right)$
and $\mathcal{A}_{U}^{j}=\mathcal{S}_{U}\left(M^{*},P^{*},\left\{ \widehat{\xi}_{k}^{j}\right\} _{k\in\mathcal{S}^{C}}\right)$
denote the outputs of Algorithm \ref{opalgo_spectrum_partitioning_stage2}
with $M^{*}$, $P^{*}$, and $\widehat{\xi}_{k}^{j};\forall k\in\mathcal{S}^{C}$
as its inputs. $\mathcal{A}_{L}^{j}$ and $\mathcal{A}_{U}^{j}$ are
the set of interested licensed and unlicensed operators corresponding
to $M^{*}$ and $P^{*}$ according to the $j^{th}$ operator. The
$j^{th}$ operator is interested in joining the market if and only
if $j\in\mathcal{A}_{L}^{j}$ or $j\in\mathcal{A}_{U}^{j}$. Let $\mathcal{S}_{L}^{true}\left(M^{*},P^{*}\right)$
and $\mathcal{S}_{U}^{true}\left(M^{*},P^{*}\right)$ denote the true
set of interested licensed operators and unlicensed operators respectively.
We have,\vspace{-1.0em}

\begin{eqnarray*}
\mathcal{S}_{L}^{true}\left(M^{*},P^{*}\right) & = & \left\{ j\in\mathcal{S}_{L}^{C}\,:\,j\in\mathcal{A}_{L}^{j}\right\} \\
\mathcal{S}_{U}^{true}\left(M^{*},P^{*}\right) & = & \left\{ j\in\mathcal{S}_{U}^{C}\,:\,j\in\mathcal{A}_{U}^{j}\right\} 
\end{eqnarray*}

To compute $\mathcal{S}_{L}^{true}\left(M^{*},P^{*}\right)$ and $\mathcal{S}_{U}^{true}\left(M^{*},P^{*}\right)$,
Algorithm \ref{opalgo_spectrum_partitioning_stage2} has to be called
$\left|\mathcal{S}_{L}^{C}\right|+\left|\mathcal{S}_{U}^{C}\right|$
times, one corresponding to each of the $\left|\mathcal{S}_{L}^{C}\right|+\left|\mathcal{S}_{U}^{C}\right|$
operators. Finally, the true value of the objective corresponding
to $M^{*}$ and $P^{*}$ is\vspace{-1.0em}

\[
U^{true}=U\left(M^{*},P^{*},\mathcal{S}_{L}^{true}\left(M^{*},P^{*}\right),\mathcal{S}_{U}^{true}\left(M^{*},P^{*}\right)\right)
\]

Please note that $U^{true}$ is implicitly a function of the tuples
$\left\{ \xi_{k}\right\} _{k\in\mathcal{S}^{C}}$ as well. While calculating
$U^{true}$, $\left\{ \xi_{k}\right\} _{k\in\mathcal{S}^{C}}$ should
be used and not any of its point estimates.

\section{Monte Carlo Integrator Design\label{sec:Monte-Carlo-Integrator}}

Algorithms \ref{opalgo_spectrum_partitioning_stage2} and \ref{opalgo_spectrum_partitioning_stage1}
rely on the computation of the objective function, $U\left(M,P,\mathcal{S}_{L},\mathcal{S}_{U}\right)$,
and the revenue function, $\mathcal{R}_{k}\left(M,P,\mathcal{S}_{L},\mathcal{S}_{U}\right)$.
In this section, we design an efficient Monte Carlo integrator to
compute these two functions. These functions are the mean of certain
random variables. Monte Carlo integrator estimates mean of a random
variable by calculating the \textit{sample mean} of the random variable.
Consider a random variable $Z\sim F_{Z}$, where $F_{Z}$ is the probability
distribution of $Z$. Let the mean and the standard deviation of $Z$
be $\mu_{Z}$ and $\sigma_{Z}$ respectively. The following recursive
formula can be used to compute the sample mean of $Z$,
\begin{equation}
\widehat{z}^{r}=\frac{\left(r-1\right)\widehat{z}^{r-1}+z_{r}}{r}\label{eq:4.0.1}
\end{equation}

\noindent where $r$ is the number of samples, $z^{r}$ is the $r^{th}$
sample of $Z$ and $\widehat{z}^{r}$ is the sample mean of $Z$ calculated
over the first $r$ samples. $\widehat{z}^{r}$ is an estimate of
$\mu_{Z}$. Note that $\widehat{z}^{r}$ itself is a random variable
with mean $\mu_{Z}$ and standard deviation $\frac{\sigma_{Z}}{\sqrt{r}}$.
According to \textit{Chebyshev's inequality}, the probability that
$\widehat{z}^{r}$ is within a $\Delta$ bound of $\mu_{Z}$ is lower
bounded as follows\vspace{-1.0em}

\begin{equation}
P\left[\left|\widehat{z}^{r}-\mu_{Z}\right|\leq\Delta\right]\geq1-\frac{\sigma_{Z}^{2}}{r\Delta^{2}}\label{eq:4.0.2}
\end{equation}

We want to design a Monte Carlo integrator whose maximum acceptable
percentage error in $\widehat{z}^{r}$ is $\beta_{1}$ with a minimum
probability of $\beta_{2}$. $\beta_{1}$ and $\beta_{2}$ captures
the ``goodness'' of estimate $\widehat{z}^{r}$; a lower $\beta_{1}$
and a higher $\beta_{2}$ implies a better estimate. To achieve this
we substitute $\Delta=\frac{\beta_{1}}{100}\mu_{Z}$ in (\ref{eq:4.0.2})
which makes the RHS of (\ref{eq:4.0.2}) equal to $1-\frac{100^{2}\sigma_{Z}^{2}}{r\beta_{1}^{2}\mu_{Z}^{2}}$.
So we have to recursively calculate $\widehat{z}^{r}$ until\vspace{-1.0em}

\begin{equation}
100^{2}\sigma_{Z}^{2}\leq r\beta_{1}^{2}\mu_{Z}^{2}\left(1-\beta_{2}\right)\label{eq:4.0.3}
\end{equation}

Inequality \ref{eq:4.0.3} can be used as one of the stopping criteria
for the Monte Carlo integrator. However, we don't know $\mu_{Z}$
and $\sigma_{Z}^{2}$ of (\ref{eq:4.0.3}); infact we want to calculate
$\mu_{Z}$. One possible heuristic would be to use the sample mean
and the sample variance in place of $\mu_{Z}$ and $\sigma_{Z}^{2}$
respectively. Sample mean can be calculated using (\ref{eq:4.0.1}).
Sample variance $\delta z^{r}$ can be computed using the following
recursive formula \cite{welford1962note}, 
\begin{equation}
\delta z^{r}=\frac{\left(r-1\right)}{r}\delta z^{r-1}+\left(r-1\right)\left(\widehat{z}^{r}-\widehat{z}^{r-1}\right)^{2}\label{eq:4.0.4}
\end{equation}

To summarize, $\mu_{Z}$ is estimated by recursively calculating the
sample mean using (\ref{eq:4.0.1}) until the following stopping criteria
is reached,\vspace{-1.0em}

\begin{equation}
100^{2}\delta z^{r}\leq r\beta_{1}^{2}\left(\widehat{z}^{r}\right)^{2}\left(1-\beta_{2}\right)\label{eq:4.0.5}
\end{equation}

Now, we discuss all the sample means which we have to calculate in
order to estimate the objective and the revenue functions. By refering
to (\ref{eq:2.4.2}), (\ref{eq:2.4.3}) and (\ref{eq:2.4.5}), we
can say that the objective function is the expected value of the net
demand served by all the interested operators in one time slot using
either licensed or opportunistic access. Let $\widehat{U}^{r}$ denote
the sample mean over $r$ samples of the net demand served by all
the interested operators in one time slot. Equation \ref{eq:2.4.7.2}
shows that the revenue function of the $k^{th}$ licensed operator
consists of two terms. The first term in (\ref{eq:2.4.7.2}) is the
expected value of the $k^{th}$ licensed operator's revenue in an
epoch generated using licensed access. Let $\widehat{R}_{k,lc}^{r}$
denote the sample mean over $r$ samples of the $k^{th}$ licensed
operator's revenue in an epoch generated using licensed access. The
second term in (\ref{eq:2.4.7.2}) is the expected value of the $k^{th}$
licensed operator's revenue in an epoch generated using opportunistic
access. This value is equal to $h_{k}\left(\mu_{k,op}^{X}\right)$
according to (\ref{eq:2.2.3}). But $\mu_{k,op}^{X}=\widetilde{\mu}_{k,op}^{x}T$
(refer to (\ref{eq:2.2.2})) and hence $h_{k}\left(\mu_{k,op}^{X}\right)=h_{k}\left(\widetilde{\mu}_{k,op}^{x}T\right)$.
$\widetilde{\mu}_{k,op}^{x}$ is the expected value of the demand
served by the $k^{th}$ operator in a time slot using opportunistic
spectrum access. Let $\widehat{U}_{k,op}^{r}$ be the estimate of
$\widetilde{\mu}_{k,op}^{x}$ over $r$ samples. Finally, $\widehat{R}_{k,lc}^{r}+h_{k}\left(\widehat{U}_{k,op}^{r}T\right)$
is the estimate of the $k^{th}$ licensed operator's revenue function
over $r$ samples. According to (\ref{eq:2.4.8}), the estimate of
the $k^{th}$ unlicensed operator's revenue function over $r$ samples
is $h_{k}\left(\widehat{U}_{k,op}^{r}T\right)$. To this end, we have
to calculate the sample means $\widehat{U}^{r}$, $\widehat{U}_{k,op}^{r}$
and $\widehat{R}_{k,lc}^{r}$ to estimate the objective and the revenue
function.

\noindent \SetInd{0.0em}{0.5em}
\SetAlgoHangIndent{0.0em}
\begin{algorithm}[t]

	\DontPrintSemicolon
	 
	\SetKwInput{Input}{Input}
	\SetKwInput{Output}{Output}

	\Input{$M$, $P$, $T$, $D$, $\phi$, $\alpha_{L}$, $\alpha_{U}$, $\mathcal{S}_{L}$, $\mathcal{S}_{U}$, and $\xi_{k};\forall k\in\mathcal{S}_{L}\bigcup\mathcal{S}_{U}$}

	\Output{$U\left(M,P,\mathcal{S}_{L},\mathcal{S}_{U}\right)$ and $\mathcal{R}_{k}\left(M,P,\mathcal{S}_{L},\mathcal{S}_{U}\right);\forall k\in\mathcal{S}_{L}\bigcup\mathcal{S}_{U}$}

	\justfy Set $\widehat{U}^{0}=0$$\;$, $\;\;\;$ $\widehat{U}^{0}_{k,op}=0\,;\,\forall k \in \mathcal{S}$$\;$, $\;$ and $\;\;\;$ $\widehat{R}^{0}_{k,lc}=0\,;\,\forall k \in \mathcal{S}_{L}$.\;

	\justfy Set $stop = False$ and $r = 0$.\;

	\justfy \While{$not\left(stop\right)$}
	{
		\justfy Set $r = r + 1$.\;

		\justfy For all $k$ in $\mathcal{S}_{L}$, sample $\theta_{k}^{r}$, $R_{k,lc}^{r}$ and $V_{k}^{r}$ from probability distribution (\ref{eq:4.2}). Set $x_{k}^{r} = \max\left(0\,,\,\theta_{k}^{r}\right)$.

		\justfy For all $k$ in $\mathcal{S}_{U}$, sample $\theta_{k}^{r}$ from the probability distribution $\mathcal{N}\left(\mu_{k}^{\theta},\left(\sigma_{k}^{\theta}\right)^{2}\right)$. Set $x_{k}^{r} = \max\left(0\,,\,\theta_{k}^{r}\right)$.

		\justfy Sort the list $\left\{ V_{k}^{r}\right\} _{k\in\mathcal{S}_{L}}$ in descending order of $V_{k}^{r}$. Let $\mathcal{T}_{1}^{r}$ be the subset of operators in $\mathcal{S}_{L}$ with the $P$ highest values of $V_{k}^{r}$. $\mathcal{T}_{1}^{r}$ are the Tier-1 operators. $\mathcal{T}_{2}^{r}=\mathcal{S}\backslash\mathcal{T}_{1}^{r}$ are the Tier-2 operators.\;

		\justfy Demand served by Tier-1 operators using licensed spectrum access are $\mbox{{\ensuremath{\widetilde{x}_{k,lc}^{r}}=\ensuremath{\min\left(x_{k}^{r},\frac{D}{M}\right)\,};\,\ensuremath{\forall}k\ensuremath{\in\mathcal{T}_{1}^{r}}}}$.\;

		\justfy Calculate modified demand, $\overline{x}_{k}^{r}$, and opportunistic channel capacity, $D_{O}^{r}$, using (\ref{eq:2.3.1.0}) and (\ref{eq:2.3.1}) respectively.\;

		\justfy Call Algorithm \ref{WaterFilling_algo} to get $\left\{ \widetilde{x}_{k,op}^{r}\right\} _{k\in\mathcal{S}}$, the demand served by operators using opportunistic spectrum access. The input to Algorithm \ref{WaterFilling_algo} are $D_{O}^{r}$ and $\left\{ \overline{x}_{k}^{r}\right\} _{k\in\mathcal{S}}$.\;

		\justfy Set $\widehat{U}^{r}=\frac{\left(r-1\right)\widehat{U}^{r-1}+\left(\underset{k\in\mathcal{T}_{1}^{r}}{\sum}\widetilde{x}_{k,lc}^{r}+\underset{k\in\mathcal{S}}{\sum}\widetilde{x}_{k,op}^{r}\right)}{r}$.\;

		\justfy Set $\widehat{R}_{k,lc}^{r}=\frac{\left(r-1\right)\widehat{R}_{k,lc}^{r-1}+R_{k,lc}^{r}}{r}$ for all $k$ in $\mathcal{T}_{1}^{r}$.\;

		\justfy Set $\widehat{R}_{k,lc}^{r}=\frac{\left(r-1\right)\widehat{R}_{k,lc}^{r-1}+0}{r}$ for all $k$ in $\mathcal{S}_{L} \backslash \mathcal{T}_{1}^{r}$.\;

		\justfy Set $\widehat{U}_{k,op}^{r}=\frac{\left(r-1\right)\widehat{U}_{k,op}^{r-1}+\widetilde{x}_{k,op}^{r}}{r}$ for all $k$ in $\mathcal{S}$.\;

		\justfy \uIf{$r=1$}
		{
			\justfy Set $\widehat{U}^{1}=0$$\;$, $\;\;\;$ $\widehat{U}^{1}_{k,op}=0\,;\,\forall k \in \mathcal{S}$$\;$, $\;$ and $\;\;\;$ $\widehat{R}^{1}_{k,lc}=0\,;\,\forall k \in \mathcal{S}_{L}$.\;
		}
		\justfy \uElse
		{
			\justfy Set $\delta U^{r}=\frac{\left(r-2\right)\delta U^{r-1}}{\left(r-1\right)}+{\scriptstyle r\left(\widehat{U}^{r}-\widehat{U}^{r-1}\right)^{2}}$ $\;\;\;\;\;\;\;\;\;\;\;$ $\delta U_{k,op}^{r}=\frac{\left(r-2\right)\delta U_{k,op}^{r-1}}{\left(r-1\right)}+{\scriptstyle r\left(\widehat{U}_{k,op}^{r}-\widehat{U}_{k,op}^{r-1}\right)^{2}}\,;\,\forall k\in\mathcal{S}$ $\delta R_{k,lc}^{r}=\frac{\left(r-2\right)\delta R_{k,lc}^{r-1}}{\left(r-1\right)}+{\scriptstyle r\left(\widehat{R}_{k,lc}^{r}-\widehat{R}_{k,lc}^{r-1}\right)^{2}}\,;\,\forall k\in\mathcal{S}_{L}$\;
		}

		\justfy Set $stop = Stop\left(r,\delta U^{r},\delta U_{k,op}^{r},\delta R_{k,lc}^{r},\widehat{U}^{r},\widehat{U}_{k,op}^{r},\widehat{R}_{k,lc}^{r}\right)$.
		
	}
		
	\justfy Set $U\left(M,P,\mathcal{S}_{L},\mathcal{S}_{U}\right)=\widehat{U}^{r}$ $\;\;\;\;\;\;\;\;\;\;\;$ $\mathcal{R}_{k}\left(M,P,\mathcal{S}_{L},\mathcal{S}_{U}\right)=\widehat{R}_{k,lc}^{r}+h_{k}\left(\widehat{U}_{k,op}^{r}T\right);\forall k\in\mathcal{S}_{L}$ $\mathcal{R}_{k}\left(M,P,\mathcal{S}_{L},\mathcal{S}_{U}\right)=h_{k}\left(\widehat{U}_{k,op}^{r}T\right)\,;\,\forall k\in\mathcal{S}_{U}$.\;
	\caption{Monte Carlo Integrator}
	\label{MonteCarloIntegrator_algo}
\end{algorithm}

\vspace{-1.0em}

We now present a proposition, which helps in generating random samples
efficiently for the Monte Carlo integrator.
\begin{prop}
\label{prop:JointProbRevDemand}Define,\vspace{-1.5em}

\begin{equation}
\varphi_{k}=\stackrel[0]{\frac{D}{M}}{\int}\vartheta^{2}f_{k}^{\theta}\left(\vartheta\right)\,d\vartheta+\frac{D}{M}\stackrel[\frac{D}{M}]{\infty}{\int}\vartheta f_{k}^{\theta}\left(\vartheta\right)\,d\vartheta-\mu_{k}^{\theta}\widetilde{\mu}_{k,lc}^{x}\label{eq:4.1}
\end{equation}

\noindent where $f_{k}^{\theta}\left(\vartheta\right)$ is the probability
density function of $\theta_{k}\left(t\right)$. Then, $\theta_{k}\left(t\right)$,
$R_{k,lc}\left(\gamma\right)$ and $V_{k}\left(\gamma\right)$ are
jointly Gaussian random variables with joint probability distribution,
\begin{equation}
\begin{bmatrix}\theta_{k}\left(t\right)\\
R_{k,lc}\left(\gamma\right)\\
V_{k}\left(\gamma\right)
\end{bmatrix}\sim\mathcal{N}\left(\psi_{k},\Sigma_{k}\right)\label{eq:4.2}
\end{equation}

\noindent for all $\gamma$ and for all $t\in\left[\left(\gamma-1\right)T+1,\gamma T\right]$
where,\vspace{-1.0em}

\begin{equation}
\psi_{k}=\begin{bmatrix}\mu_{k}^{\theta} & \mu_{k,lc}^{R} & \mu_{k,lc}^{R}\end{bmatrix}^{T}\quad\quad\quad\quad\quad\quad\quad\quad\quad\quad\label{eq:4.2.1}
\end{equation}

\vspace{-1.5em}

\begin{equation}
\Sigma_{k}=\begin{bmatrix}\left(\sigma_{k}^{\theta}\right)^{2} & \rho_{k}\frac{\sigma_{k,lc}^{R}}{\sigma_{k,lc}^{X}}\varphi_{k} & \omega_{k}\rho_{k}\frac{\sigma_{k,lc}^{R}}{\sigma_{k,lc}^{X}}\varphi_{k}\\
\rho_{k}\frac{\sigma_{k,lc}^{R}}{\sigma_{k,lc}^{X}}\varphi_{k} & \left(\sigma_{k,lc}^{R}\right)^{2} & \omega{}_{k}\left(\sigma_{k,lc}^{R}\right)^{2}\\
\omega_{k}\rho_{k}\frac{\sigma_{k,lc}^{R}}{\sigma_{k,lc}^{X}}\varphi_{k} & \omega{}_{k}\left(\sigma_{k,lc}^{R}\right)^{2} & \left(\sigma_{k,lc}^{R}\right)^{2}
\end{bmatrix}\label{eq:4.2.2}
\end{equation}
\end{prop}
\begin{IEEEproof}
Please refer to Appendix A for the proof.
\end{IEEEproof}
Recall that in (\ref{eq:4.1}), $\widetilde{\mu}_{k,lc}^{x}$ is given
by (\ref{eq:2.2.0.1}). In (\ref{eq:4.2.1}), $\mu_{k,lc}^{R}=h_{k}\left(\widetilde{\mu}_{k,lc}^{x}T\right)$
(refer to (\ref{eq:2.2.2}) and (\ref{eq:2.2.3})). In (\ref{eq:4.2.2}),
$\sigma_{k,lc}^{X}=\widetilde{\sigma}_{k,lc}^{x}\sqrt{T}$ where $\widetilde{\sigma}_{k,lc}^{x}$
is given by (\ref{eq:2.2.0.2}).

The pseudocode for the Monte Carlo integrator is given in Algorithm
\ref{MonteCarloIntegrator_algo}. The sample means $\widehat{U}^{r}$,
$\widehat{U}_{k,op}^{r}$ and $\widehat{R}_{k,lc}^{r}$ are initialized
to zero for $r=0$ (\textit{line 1}). Inside the while loop, $\widehat{U}^{r}$,
$\widehat{U}_{k,op}^{r}$ and $\widehat{R}_{k,lc}^{r}$ are computed
recursively until a stopping criteria. We discuss the stopping criteria
later in this section. In \textit{line 5}, the $r^{th}$ sample of
$\theta_{k}\left(t\right)$, $R_{k,lc}\left(\gamma\right)$ and $V_{k}\left(\gamma\right)$
are generated for all the licensed operator according to the probability
distribution given by (\ref{eq:4.2}). We have dropped the $\gamma$
and $t$ inside the paranthesis for notational simplicity. Similarly,
in \textit{line 6}, $\theta_{k}\left(t\right)$ is generated for all
the unlicensed operator. $\theta_{k}\left(t\right)$ follows the probability
distribution $\mathcal{N}\left(\mu_{k}^{\theta},\left(\sigma_{k}^{\theta}\right)^{2}\right)$
(refer to Section \ref{subsec:Demand-Model}). The $r^{th}$ sample
of $\theta_{k}\left(t\right)$, $R_{k,lc}\left(\gamma\right)$ and
$V_{k}\left(\gamma\right)$ are denoted by $\theta_{k}^{r}$, $R_{k,lc}^{r}$
and $V_{k}^{r}$ respectively. The customer demand of the $k^{th}$
operator for the $r^{th}$ sample is $x_{k}^{r}=\max\left(0,\theta_{k}^{r}\right)$.
Tier-1 and Tier-2 operators for the $r^{th}$ sample are decided in
\textit{line 7}. Licensed operators with the $P$ highest bids, $V_{k}^{r}$,
are the Tier-1 operators for the $r^{th}$ sample. $\mathcal{T}_{1}^{r}$
denotes the set of Tier-1 operators for the $r^{th}$ sample. The
remaining operators, $\mathcal{S}\backslash\mathcal{T}_{1}^{r}$,
are the Tier-2 operators for the $r^{th}$ sample. $\mathcal{T}_{2}^{r}$
denotes the set of Tier-2 operators for the $r^{th}$ sample. In \textit{lines
8-10}, demand served by the operators using licensed and opportunistic
spectrum access are calculated. Demand served by operators using licensed
and opportunistic spectrum access for the $r^{th}$ sample are denoted
using $\widetilde{x}_{k,lc}^{r}$ and $\widetilde{x}_{k,op}^{r}$
respectively.

The sample means $\widehat{U}^{r}$, $\widehat{U}_{k,op}^{r}$ and
$\widehat{R}_{k,lc}^{r}$ are calculated in \textit{lines 11-14} using
recursive formulas analogous to (\ref{eq:4.0.1}). The formula to
update $\widehat{U}^{r}$ is shown in \textit{line 11}. The term $\underset{k\in\mathcal{T}_{1}^{r}}{\sum}\widetilde{x}_{k,lc}^{r}+\underset{k\in\mathcal{S}}{\sum}\widetilde{x}_{k,op}^{r}$
is the net demand served by all the operators in a time slot for the
$r^{th}$ sample. The formula to update $\widehat{R}_{k,lc}^{r}$
is shown in \textit{lines 12 and 13}. If the $k^{th}$ licensed operator
is a Tier-1 operator for the $r^{th}$ sample, then it earns a revenue
of $R_{k,lc}^{r}$ in an epoch using licensed spectrum access (\textit{line
12}). But if the $k^{th}$ licensed operator is a Tier-2 operator
for the $r^{th}$ sample, then it earns a revenue of $0$ using licensed
spectrum access (\textit{line 13}). $\widehat{U}_{k,op}^{r}$ is updated
in \textit{line 14}. The operators serves $\widetilde{x}_{k,op}^{r}$
customer demand using opportunistic spectrum access (\textit{line
14}). The sample variance corresponding to sample means $\widehat{U}^{r}$,
$\widehat{U}_{k,op}^{r}$ and $\widehat{R}_{k,lc}^{r}$ are calculated
in \textit{lines 15-18}. These variances are initialized to zero for
the $1^{st}$ sample (\textit{line 16}) and updated using recursive
formulas similar to (\ref{eq:4.0.4}) for $r>1$ (\textit{line 18}).
In \textit{line 19}, the $Stop\left(\cdot\right)$ function decides
whether to stop the Monte Carlo integrator. The stopping criteria
is based on (\ref{eq:4.0.5}). The $Stop\left(\cdot\right)$ function
returns $True$ if and only if all the following conditions are met,\vspace{-1.2em}

\begin{equation}
100^{2}\delta U^{r}\leq r\beta_{1}^{2}\left(\widehat{U}^{r}\right)^{2}\left(1-\beta_{2}\right)\qquad\qquad\qquad\quad\label{eq:4.2.3}
\end{equation}

\vspace{-1.0em}

\begin{equation}
100^{2}\delta U_{k,op}^{r}\leq r\beta_{1}^{2}\left(\widehat{U}_{k,op}^{r}\right)^{2}\left(1-\beta_{2}\right)\,;\,\forall k\in\mathcal{S}\label{eq:4.2.4}
\end{equation}

\vspace{-1.0em}

\begin{equation}
100^{2}\delta R_{k,lc}^{r}\leq r\beta_{1}^{2}\left(\widehat{R}_{k,lc}^{r}\right)^{2}\left(1-\beta_{2}\right)\,;\,\forall k\in\mathcal{S}_{L}\qquad\quad\label{eq:4.2.5}
\end{equation}

\vspace{-1.0em}

\begin{equation}
r\geq r_{min}\qquad\qquad\qquad\qquad\qquad\qquad\label{eq:4.2.6}
\end{equation}

The last condition $r\geq r_{min}$ ensures that the Monte Carlo integrator
samples the mean over atleast $r_{min}$ samples. We have used $r_{min}=10000$,
$\beta_{1}=1$ and $\beta_{2}=0.99$ unless states otherwise. Finally,
the estimated values of the objective function and the revenue function
are set in \textit{line 20} according to what we have discussed before
in this section (refer to the paragraph before Proposition \ref{prop:JointProbRevDemand}).

\section{Numerical Results\label{sec:Numerical-Results}}

In this section, we conduct numerical simulations to benchmark the
algorithms developed in the previous sections. We also explore how
the optimal solution $M^{*}$ and $P^{*}$ varies with interference
parameters. Throughout this section, each time slot has a duration
of one week and lease duration of licensed channels is one year. Hence,
$T=52$. In all our simulations we have: \textit{(i)} $h_{k}\left(\mu_{k,a}^{X}\right)=a_{k}\mu_{k,a}^{X}$
where $a_{k}>0$. \textit{(ii)} $\sigma_{k,a}^{R}=\eta_{k}h_{k}\left(\mu_{k,a}^{X}\right)$
where $\eta_{k}>0$ is the coefficient of variation of $R_{k,a}\left(\gamma\right)$.
\textit{(iii)} $\lambda_{k}=\Lambda_{k}\cdot\left(a_{k}\mu_{k}^{\theta}T\right)$
where $\Lambda_{k}\in\left[0,1\right]$ and the term $a_{k}\mu_{k}^{\theta}T$
is the mean revenue of the $k^{th}$ operator in an epoch if it can
serve all its customer demand in every time slot. \textit{(iv)} The
maximum capacity of the entire bandwidth $D$ is a fraction $\upsilon$
of the sum of $\mu_{k}^{\theta}$ of all the candidate operators,
i.e. $D=\upsilon\underset{k\in\mathcal{S}^{C}}{\sum}\mu_{k}^{\theta}$
where $\mathcal{S}^{C}=\mathcal{S}_{L}^{C}\bigcup\mathcal{S}_{U}^{C}$.
Given our choice of $h_{k}\left(\mu_{k,a}^{X}\right)$, $\sigma_{k,a}^{R}$,
and $\lambda_{k}$, the tuple $\xi_{k}$ is equivalent to $\left(\mu_{k}^{\theta},\,\sigma_{k}^{\theta},\,a_{k},\,\eta_{k},\,\rho_{k},\,\omega_{k},\,\Lambda_{k}\right)$
in this section.

\noindent 
\begin{figure}[t]
\begin{centering}
\includegraphics[scale=0.67]{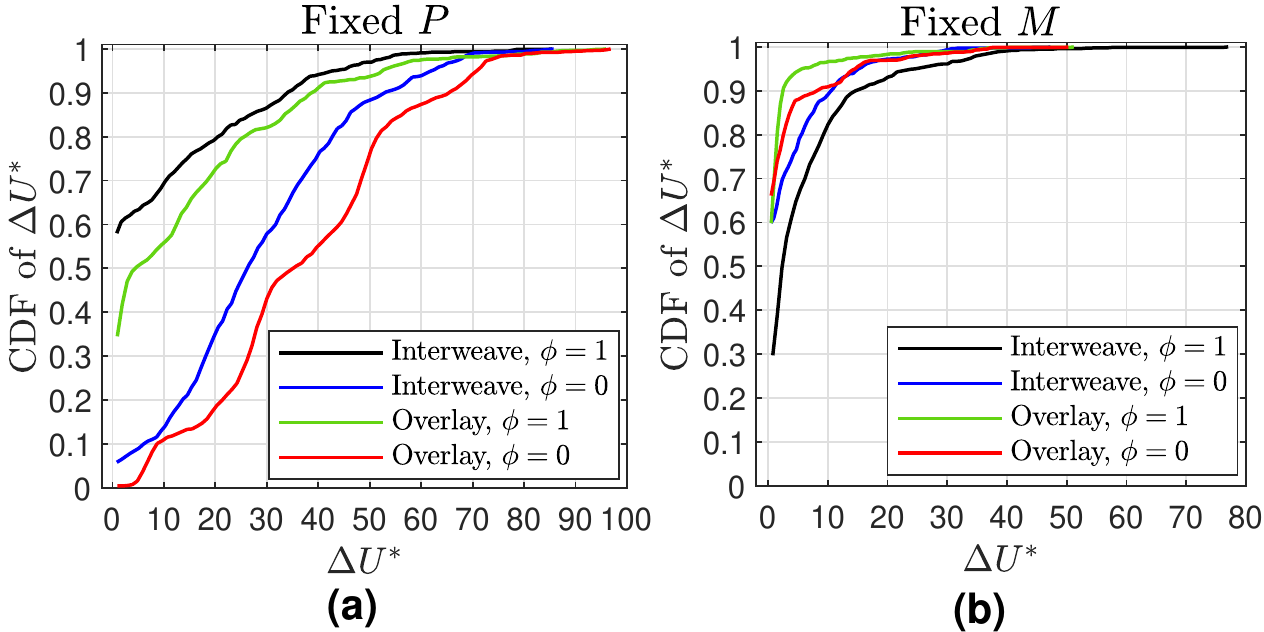}
\par\end{centering}
\caption{Cumulative distribution function of the percentage increase in objective
function, $\Delta U^{*}$, for four different types of opportunistic
spectrum access when the sub-optimal algorithm is: (a) optimizing
$M$ while holding $P$ fixed. (b) optimizing $P$ while holding $M$
fixed.\vspace{-1.0em}}
\label{sim_benchmark}
\end{figure}

\vspace{-1.0em}

\subsection{Benefit of joint optimization of $M$ and $P$\label{subsec:Benefit-of-joint}}

In our first numerical simulation, we analyze the increase in spectrum
utilization that one can obtain using joint optimization of $M$ and
$P$ when compared to optimizing $M$ while holding $P$ fixed and
vice-versa. Our numerical setup is as follows. There are four candidate
licensed operators and no candidate unlicensed operator. There are
$10$ parameters which completely defines a market setting: $\mu_{k}^{\theta}$,
$\sigma_{k}^{\theta}$, $a_{k}$, $\eta_{k}$, $\rho_{k}$, $\omega_{k}$,
$\Lambda_{k}$, $\upsilon$, $\alpha_{L}$, and $\alpha_{U}$. We
generate $1000$ such market settings by randomly selecting these
$10$ parameters from uniform distributions each of which is associated
with a certain range. The range of the parameters $\mu_{k}^{\theta}$,
$\sigma_{k}^{\theta}$, $a_{k}$, $\eta_{k}$, $\rho_{k}$, $\omega_{k}$,
and $\Lambda_{k}$ for all the operators are $\left[0.75,1.0\right]$,
$\left[0.25,0.75\right]$, $\left[0.9,1.1\right]$, $\left[0.25,0.75\right]$,
$\left[0.5,0.9\right]$, $\left[0.85,0.95\right]$, and $\left[0.25,1.0\right]$
respectively. The range of $\upsilon$, $\alpha_{L}$, and $\alpha_{U}$
are $\left[0.5,1.0\right]$, $\left[0.75,1.0\right]$, and $\left[0.75,1.0\right]$
respectively. While generating $\alpha_{L}$ and $\alpha_{U}$, we
ensure that $\alpha_{L}\le\alpha_{U}$.

The optimal value of the objective function corresponding to Algorithm
\ref{opalgo_spectrum_partitioning_stage1} is $U^{*}$. We compare
Algorithm \ref{opalgo_spectrum_partitioning_stage1} with a sub-optimal
algorithm. Let the optimal value of the objective function corresponding
to a sub-optimal algorithm be $\widehat{U}^{*}$. The percentage increase
in the objective function is $\Delta U^{*}=\frac{U^{*}-\widehat{U}^{*}}{D}\cdot100$.
The reason for having $D$ in the denominator is as follows. The objective
function given by (\ref{eq:2.4.5}) is the mean demand served by all
the operators in one time slot which cannot be greater than $D$,
the maximum capacity of the entire bandwidth. Hence, $U^{*},\widehat{U}^{*}\leq D$
which implies that $U^{*}-\widehat{U}^{*}\leq D$. We compute $\Delta U^{*}$
for sub-optimal algorithms and plot the cumulative distribution function
(CDF) of $\Delta U^{*}$ in Figure \ref{sim_benchmark}. Recall that
$\phi$ can be $0$ or $1$, and the OSA strategy can be either interweave
or overlay. So there are four possible combination of OSA. For a given
sub-optimal algorithm, we compute CDFs for all the four combinations.

We consider two sub-optimal algorithms. For the first algorithm, $P$
is fixed and $M$ is optimized. An intuitive choice of $P$ is the
number of candidate licensed operators. In that way, every candidate
licensed operators wins a licensed channel in every epoch. For the
second algorithm, $M$ is fixed and $P$ is optimized. We set $M=\left\lfloor \frac{D}{\vartheta}\right\rfloor $
where $\left\lfloor \cdot\right\rfloor $ is the floor function and
$\vartheta=\frac{1}{\left|\mathcal{S}^{C}\right|}\underset{k\in\mathcal{S}^{C}}{\sum}\mu_{k}^{\theta}$
is the sample mean of the mean of an operator's customer demand. This
choice of $M$ is to ensure that the bandwidth $\vartheta$ of a licensed
channel is neither too high that most of it is wasted and neither
too low that a licensed operator has to reject most of its cushtomer
demand.

In Figure \ref{sim_benchmark}, a lower value of CDF for a given $\Delta U^{*}$
implies that the difference in spectrum utilization between joint
optimization and the sub-optimal algorithm is higher. By comparing
Figures \ref{sim_benchmark}.a and \ref{sim_benchmark}.b we can say
that joint optimization leads to more improvement in spectrum utilization
when $P$ is fixed rather than when $M$ is fixed. Based on Figure
\ref{sim_benchmark}.a, we can say that when $P$ is fixed, joint
optimization leads to more improvement in spectrum utilization for:
(i) overlay strategy than interweave strategy when $\phi$ is fixed.
(ii) $\phi=0$ than $\phi=1$. Based on Figure \ref{sim_benchmark}.b,
we can say that when $M$ is fixed, joint optimization leads to more
improvement in spectrum utilization for interweave strategy than overlay
strategy when $\phi$ is fixed. We don't observe any such systematic
trend for $\phi$ when $M$ is fixed.\vspace{-2.0em}

\noindent 
\begin{figure}[t]
\begin{centering}
\includegraphics[scale=0.8]{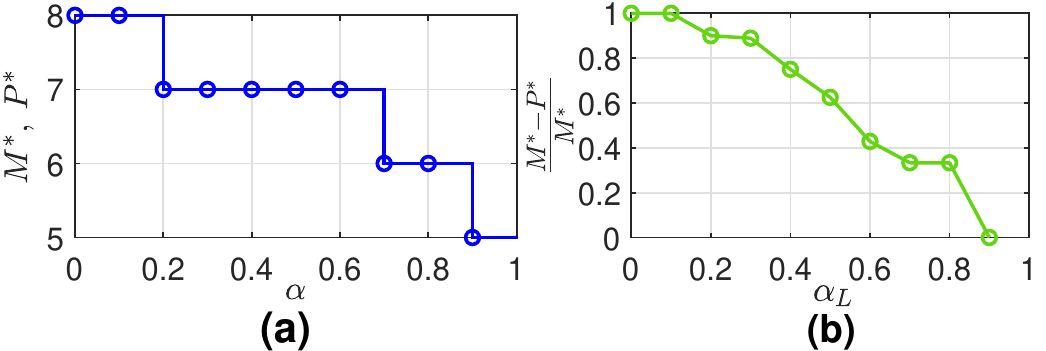}
\par\end{centering}
\caption{(a) Plots showing the effect of interference parameter $\alpha$ for
a market containing only candidate licensed operators on the optimal
number of channel, $M^{*}$, and the optimal number of licensed channels,
$P^{*}$. (b) Plots showing the effect of interference parameter of
licensed channel $\alpha_{L}$ for a market containing both candidate
licensed operators and unlicensed operators on the ratio of the bandwidth
allocated for unlicensed channels, $\frac{M^{*}-P^{*}}{M^{*}}$.\vspace{-1.0em}}
\label{sim_interference}
\end{figure}

\subsection{Effect of interference parameters}

Our second numerical simulation is to study the effect of interference
parameter on optimal solution. We consider two simulation setups.
The first simulation setup is as follows. For this setup, $\alpha_{L}=\alpha_{U}=\alpha$.
There are $8$ candidate licensed operators and no candidate unlicensed
operators. We consider a homogeneous market setting. The minimum revenue
requirement $\lambda_{k}$ is set to zero for all the operators which
ensures that all the operators join the market. The remaining parameters
of the market are: $\mu_{k}^{\theta}=1$, $\sigma_{k}^{\theta}=0.5$,
$a_{k}=1$, $\eta_{k}=0.5$, $\rho_{k}=0.8$, and $\omega_{k}=0.9$
for all $k$'s. Also, $\upsilon=0.8$. We study how $M^{*}$ and $P^{*}$
varies with $\alpha$. The simulation result is shown in Figure \ref{sim_interference}.a.
Since there are no candidate unlicensed operators, it is intuitive
that there are no unlicensed channels, i.e. $M^{*}=P^{*}$. Figure
\ref{sim_interference}.a shows that $M^{*}$ decreases with increase
in $\alpha$. This can be explained as follows. If $M$ is low, the
bandwidth, and hence the capacity of each licensed channel is high.
Therefore, a licensed operator can serve more customer demand using
the allocated licensed channel thereby increasing spectrum utilization.
But if $M$ is too low, only few of the $8$ licensed operators are
allocated the licensed channels in an epoch. The remaining operators
who uses channels opportunistically as Tier-2 operators. The efficiency
of opportunistic access is decided by $\alpha$. If $\alpha$ is low,
it is better to have fewer Tier-2 operators in an epoch because opportunistic
spectrum access is inefficient. This can be ensured with a higher
$M$ so that there are more Tier-1 operators in every epoch.

In our second simulation setup, we include candidate unlicensed operators.
The simulation setup is similar to the first setup but differs in
the following ways. \textit{First}, out of the $8$ operators, four
are candidate licensed operators and four are candidate unlicensed
operators. \textit{Second}, the interference parameters $\alpha_{L}$
and $\alpha_{U}$ are not same. We set $\alpha_{U}=0.9$ and vary
$\alpha_{L}$ from $0$ to $0.9$. We study how the ratio of the bandwidth
allocated for unlicensed channels characterized by the ratio $\frac{M^{*}-P^{*}}{M^{*}}$
changes with $\alpha_{L}$. This is shown in Figure \ref{sim_interference}.b.
Unlike the previous simulation setup, the current simulation setup
has candidate unlicensed operators. Therefore, we expect that there
will be unlicensed channels dedicated for the candidate unlicensed
operators. But the question is: what portion of the bandwidth should
be allocated for unlicensed channels? If $\alpha_{L}$ is high, most
of the bandwidth can be reserved for licensed channels because even
if the Tier-1 operators are not using the licesnsed channels, the
Tier-2 operators can use the remaining capacity of the licensed channels
efficiently. But as $\alpha_{L}$ decreases, the opportunistic access
of licensed channels becomes inefficient. Therefore, it is better
to reserve higher portion of the bandwidth for unlicensed channels.

\subsection{Market competition vs Spectrum Utilization}

For most markets, an increase in competition improves the social welfare.
In our setup, we use the number of interested operators, $\left|\mathcal{S}_{L}\right|+\left|\mathcal{S}_{U}\right|$,
as the measure of market competition and spectrum utilization as the
measure of social welfare. In this numerical simulation, we show that
there exists market setups where an increase in $\left|\mathcal{S}_{L}\right|+\left|\mathcal{S}_{U}\right|$
decreases spectrum utilization. The simulation setup and the definition
of $\Delta U^{*}$ is similar to Section \ref{subsec:Benefit-of-joint}
but differs in the following ways.\textit{ First,} in this setup we
have three candidate licensed operators and three candidate unlicensed
operators. \textit{Second,} the sub-optimal algorithm in this setup
finds $M$ and $P$ that maximize $\left|\mathcal{S}_{L}\right|+\left|\mathcal{S}_{U}\right|$
instead of the objective function defined in equation (17). If there
are multiple values of $M$ and $P$ that maximize $\left|\mathcal{S}_{L}\right|+\left|\mathcal{S}_{U}\right|$,
we choose the ones that maximize the objective function defined in
equation (17).

\noindent 
\begin{figure}[t]
\begin{centering}
\includegraphics[scale=0.65]{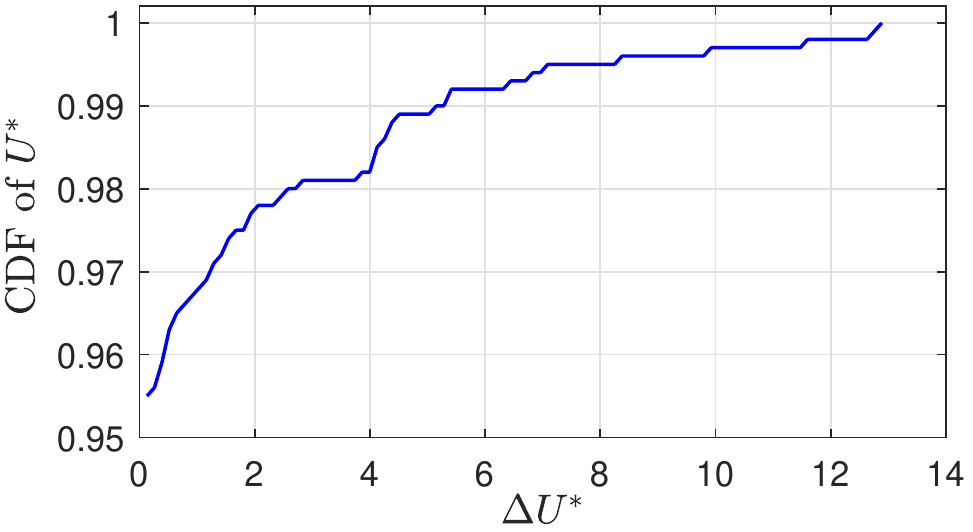}
\par\end{centering}
\caption{Cumulative distribution function of the percentage increase in objective
function, $\Delta U^{*}$, when the sub-optimal algorithm is to choose
the value of $M$ and $P$ that maximize the number of interested
operators.\vspace{-1.0em}}
\label{sim_competition}
\end{figure}

\vspace{-1.0em}

The simulation result is shown in Figure \ref{sim_competition} where
we plot the CDF of $\Delta U^{*}$ for $1000$ market setups. To establish
our claim that maximizing $\left|\mathcal{S}_{L}\right|+\left|\mathcal{S}_{U}\right|$
doesn't necessarily maximize spectrum utilization, we want to find
market setups where $\Delta U^{*}$ is strictly greater than $0$.
We can see that for $\left(1-0.955\right)\cdot100\%=4.5\%$ of the
market setups, $\Delta U^{*}>0$. This establishes our claim that
there are market setups, however few, where maximizing $\left|\mathcal{S}_{L}\right|+\left|\mathcal{S}_{U}\right|$
doesn't necessarily maximize spectrum utilization. However, for these
$4.5\%$ of the market setups, $\Delta U^{*}$ is upper bounded by
$13\%$ implying only a marginal improvement in spectrum utilization.

\section{Conclusion\label{sec:Conclusion}}

In this paper, we designed an optimization algorithm to partition
a bandwidth into channels and further decide the number of licensed
channels in order to maximize spectrum utilization. The access to
this bandwidth is governed by a tiered spectrum access model inspired
by the CBRS band. We first propose a system model which accurately
captures various aspects of the tiered spectrum access model. Based
on this model, we formulate our optimization problem as a two-staged
Stackelberg game and then designed algorithms to solve the Stackelberg
game. Finally, we get numerical results to benchmark our algorithm
and to also study certain optimal trends of spectrum partitioning
and licensing as a function of interference parameters.

There can be various directions for future research related to generalization
of the Stackelberg Game model. \textit{First}, is to capture collusion
between operators in Stage-2 of the Stackelberg Game.\textit{ Second},
in our current model, every operator is assumed to be equally pessimistic.
It would be interesting to associate each operator with a degree of
pessimism.

\appendices{}

\section{Proof of Proposition 2}

Throughout this derivation, $t$ represents the $t^{th}$ time slot
of $\gamma^{th}$ epoch, i.e. $t\in\left[\left(\gamma-1\right)T+1,\gamma T\right]$.
We start by deriving the stochastic model between $\theta_{k}\left(t\right)$
and $X_{k,lc}\left(\gamma\right)$, the net demand served by the $k^{th}$
operator in $\gamma^{th}$ epoch using licensed spectrum access. For
Gaussian random variables $\theta_{k}\left(t\right)$ and $X_{k,lc}\left(\gamma\right)$,
their joint Gaussian distribution is characterised by their mean and
their covariance matrix. The mean and the variance of $\theta_{k}\left(t\right)$
are $\mu_{k}^{\theta}$ and $\left(\sigma_{k}^{\theta}\right)^{2}$
respectively. The mean and the variance of $X_{k,lc}\left(\gamma\right)$
are $\mu_{k,lc}^{X}$ and $\left(\sigma_{k,lc}^{X}\right)^{2}$ respectively.
So, to find the joint Gaussian distribution of $\theta_{k}\left(t\right)$
and $X_{k,lc}\left(\gamma\right)$, all we have to derive is the covariance
of $\theta_{k}\left(t\right)$ and $X_{k,lc}\left(\gamma\right)$.
To do so, we are going to first recall few notations from Section
II-B. We have, $x_{k}\left(t\right)=\max\left(0,\theta_{k}\left(t\right)\right)$,
$\widetilde{x}_{k,lc}\left(t\right)=\min\left(x_{k}\left(t\right),\frac{D}{M}\right)$
and $X_{k,lc}\left(\gamma\right)=\stackrel[t=\left(\gamma-1\right)T+1]{\gamma T}{\sum}\widetilde{x}_{k,lc}\left(t\right)$.
Define,

\begin{equation}
\overline{X}_{k,lc}\left(\gamma,t\right)=\stackrel[v=\left(\gamma-1\right)T+1]{t-1}{\sum}\widetilde{x}_{k,lc}\left(v\right)+\stackrel[v=t+1]{\gamma T}{\sum}\widetilde{x}_{k,lc}\left(v\right)\label{eq:7.1}
\end{equation}

\noindent where, $\widetilde{x}_{k,lc}\left(v\right)=\min\left(\max\left(0,\theta_{k}\left(v\right)\right),\frac{D}{M}\right)$.
The mean of $\overline{X}_{k,lc}\left(\gamma,t\right)$ is $E\left[\overline{X}_{k,lc}\left(\gamma,t\right)\right]=\widetilde{\mu}_{k,lc}^{x}\left(T-1\right)$.
The covariance of $\theta_{k}\left(t\right)$ and $X_{k,lc}\left(\gamma\right)$
is\vspace{-1.0em}

\begin{eqnarray}
\varphi_{k} & = & E\left[\left(\theta_{k}\left(t\right)-\mu_{k}^{\theta}\right)\left(X_{k,lc}\left(\gamma\right)-\mu_{k,lc}^{X}\right)\right]\nonumber \\
 & = & E\left[\left(\theta_{k}\left(t\right)-\mu_{k}^{\theta}\right)\left(\overline{X}_{k,lc}\left(\gamma,t\right)-\widetilde{\mu}_{k,lc}^{x}\left(T-1\right)\right)\right.\nonumber \\
 &  & \left.\quad+\left(\theta_{k}\left(t\right)-\mu_{k}^{\theta}\right)\left(\widetilde{x}_{k,lc}\left(t\right)-\widetilde{\mu}_{k,lc}^{x}\right)\right]\nonumber \\
 & = & E\left[\left(\theta_{k}\left(t\right)-\mu_{k}^{\theta}\right)\left(\overline{X}_{k,lc}\left(\gamma,t\right)-\widetilde{\mu}_{k,lc}^{x}\left(T-1\right)\right)\right]\nonumber \\
 &  & +E\left[\left(\theta_{k}\left(t\right)-\mu_{k}^{\theta}\right)\left(\widetilde{x}_{k,lc}\left(t\right)-\widetilde{\mu}_{k,lc}^{x}\right)\right]\label{eq:7.2}
\end{eqnarray}

In (\ref{eq:7.2}), $\theta_{k}\left(t\right)$ and $\overline{X}_{k,lc}\left(\gamma,t\right)$
are independent random variables. This is because $\theta_{k}\left(t\right)$
are iid random variables and according to (\ref{eq:7.1}), $\overline{X}_{k,lc}\left(\gamma,t\right)$
is not directly dependent on $\theta_{k}\left(t\right)$. Therefore,
the first term of (\ref{eq:7.2}) is zero. We have,\vspace{-1.0em}

\begin{eqnarray*}
\varphi_{k} & = & E\left[\left(\theta_{k}\left(t\right)-\mu_{k}^{\theta}\right)\left(\widetilde{x}_{k,lc}\left(t\right)-\widetilde{\mu}_{k,lc}^{x}\right)\right]\\
 & = & E\left[\theta_{k}\left(t\right)\widetilde{x}_{k,lc}\left(t\right)\right]-\mu_{k}^{\theta}\widetilde{\mu}_{k,lc}^{x}\\
 & = & E\left[\theta_{k}\left(t\right)\min\left(\max\left(0,\theta_{k}\left(t\right)\right),\frac{D}{M}\right)\right]-\mu_{k}^{\theta}\widetilde{\mu}_{k,lc}^{x}
\end{eqnarray*}
\vspace{-2.0em}

\begin{equation}
\qquad=\stackrel[0]{\frac{D}{M}}{\int}\vartheta^{2}f_{k}^{\theta}\left(\vartheta\right)\,d\vartheta+\frac{D}{M}\stackrel[\frac{D}{M}]{\infty}{\int}\vartheta f_{k}^{\theta}\left(\vartheta\right)\,d\vartheta-\mu_{k}^{\theta}\widetilde{\mu}_{k,lc}^{x}\label{eq:7.3}
\end{equation}

\noindent where, $f_{k}^{\theta}\left(\vartheta\right)$ is the probability
density function of $\theta_{k}\left(t\right)$. To this end, we have
derived the joint Gaussian distribution of $\theta_{k}\left(t\right)$
and $X_{k,lc}\left(\gamma\right)$. We have,\vspace{-1.1em}

\begin{equation}
\begin{bmatrix}\theta_{k}\left(t\right)\\
X_{k,lc}\left(\gamma\right)
\end{bmatrix}\sim\mathcal{N}\left(\begin{bmatrix}\mu_{k}^{\theta}\\
\mu_{k,lc}^{X}
\end{bmatrix},\begin{bmatrix}\left(\sigma_{k}^{\theta}\right)^{2} & \varphi_{k}\\
\varphi_{k} & \left(\sigma_{k,lc}^{X}\right)^{2}
\end{bmatrix}\right)\label{eq:7.4}
\end{equation}

The joint Gaussian distribution of $\theta_{k}\left(t\right)$ and
$X_{k,lc}\left(\gamma\right)$ is given by (\ref{eq:7.4}), that of
$R_{k,lc}\left(\gamma\right)$ and $X_{k,lc}\left(\gamma\right)$
is given by (7) and that of $R_{k,lc}\left(\gamma\right)$ and $V_{k}\left(\gamma\right)$
is given by (10). We want to derive the joint Gaussian distribution
of $\theta_{k}\left(t\right)$, $R_{k,lc}\left(\gamma\right)$ and
$V_{k}\left(\gamma\right)$. Just like $\theta_{k}\left(t\right)$
and $X_{k,lc}\left(\gamma\right)$, all we need to know to derive
the joint Gaussian distribution of $\theta_{k}\left(t\right)$, $R_{k,lc}\left(\gamma\right)$
and $V_{k}\left(\gamma\right)$ are their mean and covariance matrix.
We have already discussed the mean and variance of $\theta_{k}\left(t\right)$
in the beginning of this section. The mean and variance of both $R_{k,lc}\left(\gamma\right)$
and $V_{k}\left(\gamma\right)$ are $\mu_{k,lc}^{R}$ and $\left(\sigma_{k,lc}^{R}\right)^{2}$
respectively (refer to (10)). According to (10), the covariance of
$R_{k,lc}\left(\gamma\right)$ and $V_{k}\left(\gamma\right)$ is
$\omega{}_{k}\left(\sigma_{k,lc}^{R}\right)^{2}$. All we have to
derive is the covariance of two set of random variables. First, $\theta_{k}\left(t\right)$
and $R_{k,lc}\left(\gamma\right)$ and second, $\theta_{k}\left(t\right)$
and $V_{k}\left(\gamma\right)$. The following proposition is important
for this derivation.
\begin{prop}
\label{prop:ConditionalGaussian}Consider jointly Gaussian variables
$A$ and $B$,\vspace{-0.5em}

\begin{equation}
\begin{bmatrix}A\\
B
\end{bmatrix}\sim\mathcal{N}\left(\begin{bmatrix}\mu_{A}\\
\mu_{B}
\end{bmatrix},\begin{bmatrix}\sigma_{A}^{2} & \sigma_{AB}\\
\sigma_{AB} & \sigma_{B}^{2}
\end{bmatrix}\right)\label{eq:7.5}
\end{equation}

\noindent where $\mu_{A}$ and $\sigma_{A}$ are the mean and standard
deviation of $A$ respectively, $\mu_{B}$ and $\sigma_{B}$ are the
mean and standard deviation of $B$ respectively and $\rho_{AB}$
is the correlation coefficient of $A$ and $B$. Let $\mathcal{L}\left(A\,|\,B=b\right)$
denote the conditional distribution of $A$ given $B=b$. Then, $\mathcal{L}\left(A\,|\,B=b\right)$
is a Gaussian distribution with mean $\mu_{A}+\frac{\sigma_{AB}}{\sigma_{B}^{2}}\left(b-\mu_{B}\right)$
and standard deviation $\sqrt{\sigma_{A}^{2}-\frac{\sigma_{AB}^{2}}{\sigma_{B}^{2}}}$.
Mathematically,\vspace{-1.0em}

\begin{equation}
{\textstyle \mathcal{L}\left(A\,|\,B=b\right)=}\mathcal{N}\left(\mu_{A}+\frac{\sigma_{AB}}{\sigma_{B}^{2}}\left(b-\mu_{B}\right),\left(\sigma_{A}^{2}-\frac{\sigma_{AB}^{2}}{\sigma_{B}^{2}}\right)\right)\label{eq:7.6}
\end{equation}
\end{prop}
\begin{IEEEproof}
The proof of a generalized version of this proposition can be found
in \cite[Chapter 3]{eaton1983multivariate}.
\end{IEEEproof}
The covariance of $\theta_{k}\left(t\right)$ and $R_{k,lc}\left(\gamma\right)$
is\vspace{-1.0em}

\begin{equation}
\overline{\varphi}_{k}=E\left[\left(\theta_{k}\left(t\right)-\mu_{k}^{\theta}\right)\left(R_{k,lc}\left(\gamma\right)-\mu_{k,lc}^{R}\right)\right]\label{eq:7.7}
\end{equation}

Let $f_{k}^{X}\left(\phi\right)$ be the probability density function
of $X_{k,lc}\left(\gamma\right)$. Define, 
\begin{equation}
{\textstyle h\left(\phi\right)=E\left[\left(\theta_{k}\left(t\right)-\mu_{k}^{\theta}\right)\left(R_{k,lc}\left(\gamma\right)-\mu_{k,lc}^{R}\right)\mid X_{k,lc}\left(\gamma\right)=\phi\right]}\label{eq:7.8}
\end{equation}
where $h\left(\phi\right)$ is the covariance of $\theta_{k}\left(t\right)$
and $R_{k,lc}\left(\gamma\right)$ conditioned on $X_{k,lc}\left(\gamma\right)=\phi$.
Using the \textit{law of total expectation}, (\ref{eq:7.7}) can be
equivalently written as\vspace{-1.2em}

\begin{eqnarray}
\overline{\varphi}_{k} & = & \stackrel[-\infty]{\infty}{\int}h\left(\phi\right)f_{k}^{X}\left(\phi\right)\,d\phi\label{eq:7.9}
\end{eqnarray}

\noindent Based on (\ref{eq:7.4}) and (7), if $X_{k,lc}\left(\gamma\right)$
is given, $R_{k,lc}\left(\gamma\right)$ and $\theta_{k}\left(t\right)$
are independent of each other. Therefore, (\ref{eq:7.8}) can be equivalently
written as\vspace{-1.0em}

\begin{eqnarray}
h\left(\phi\right) & = & E\left[\left(\theta_{k}\left(t\right)-\mu_{k}^{\theta}\right)\mid X_{k,lc}\left(\gamma\right)=\phi\right]\nonumber \\
 &  & \cdot E\left[\left(R_{k,lc}\left(\gamma\right)-\mu_{k,lc}^{R}\right)\mid X_{k,lc}\left(\gamma\right)=\phi\right]\nonumber \\
 & = & \left(E\left[\theta_{k}\left(t\right)\mid X_{k,lc}\left(\gamma\right)=\phi\right]-\mu_{k}^{\theta}\right)\nonumber \\
 &  & \cdot\left(E\left[R_{k,lc}\left(\gamma\right)\mid X_{k,lc}\left(\gamma\right)=\phi\right]-\mu_{k,lc}^{R}\right)\label{eq:7.10}
\end{eqnarray}

Using (\ref{eq:7.4}) and Proposition \ref{prop:ConditionalGaussian},\vspace{-1.2em}

\begin{equation}
E\left[\theta_{k}\left(t\right)\mid X_{k,lc}\left(\gamma\right)=\phi\right]=\mu_{k}^{\theta}+\frac{\varphi_{k}}{\left(\sigma_{k,lc}^{X}\right)^{2}}\left(\phi-\mu_{k,lc}^{X}\right)\label{eq:7.11}
\end{equation}

Using (7) and Proposition \ref{prop:ConditionalGaussian},\vspace{-1.2em}

\begin{equation}
E\left[R_{k,lc}\left(\gamma\right)\mid X_{k,lc}\left(\gamma\right)=\phi\right]=\mu_{k,lc}^{R}+\frac{\rho_{k}\sigma_{k,lc}^{X}\sigma_{k,lc}^{R}}{\left(\sigma_{k,lc}^{X}\right)^{2}}\left(\phi-\mu_{k,lc}^{X}\right)\label{eq:7.12}
\end{equation}

Substituting (\ref{eq:7.10}), (\ref{eq:7.11}) and (\ref{eq:7.12})
in (\ref{eq:7.9}) we get,\vspace{-1.2em}

\begin{eqnarray}
\overline{\varphi}_{k} & = & \frac{\rho_{k}\sigma_{k,lc}^{R}\varphi_{k}}{\left(\sigma_{k,lc}^{X}\right)^{3}}\stackrel[-\infty]{\infty}{\int}\left(\phi-\mu_{k,lc}^{X}\right)^{2}f_{k}^{X}\left(\phi\right)\,d\phi\label{eq:7.13}
\end{eqnarray}

The integral in (\ref{eq:7.13}) is the variance of $X_{k,lc}\left(\gamma\right)$
which is equal to $\left(\sigma_{k,lc}^{X}\right)^{2}$. Therefore,
(\ref{eq:7.13}) is equal to\vspace{-0.75em}

\begin{equation}
\overline{\varphi}_{k}=\frac{\rho_{k}\sigma_{k,lc}^{R}\varphi_{k}}{\left(\sigma_{k,lc}^{X}\right)^{3}}\left(\sigma_{k,lc}^{X}\right)^{2}=\rho_{k}\frac{\sigma_{k,lc}^{R}}{\sigma_{k,lc}^{X}}\varphi_{k}\label{eq:7.14}
\end{equation}

Now, we have to derive the covariance of $\theta_{k}\left(t\right)$
and $V_{k}\left(\gamma\right)$. This derivation is same as the derivation
for the covariance of $\theta_{k}\left(t\right)$ and $R_{k,lc}\left(\gamma\right)$
and has been skipped for brevity. We simply state that the covariance
of $\theta_{k}\left(t\right)$ and $V_{k}\left(\gamma\right)$ is
$\omega_{k}\rho_{k}\frac{\sigma_{k,lc}^{R}}{\sigma_{k,lc}^{X}}\varphi_{k}$.
Finally, based on our derivation, the mean, $\psi_{k}$, and the covariance
matrix, $\Sigma_{k}$, of $\theta_{k}\left(t\right)$, $R_{k,lc}\left(\gamma\right)$
and $V_{k}\left(\gamma\right)$ are given by (32) and (33) respectively.
Hence, the joint probability distribution of $\theta_{k}\left(t\right)$,
$R_{k,lc}\left(\gamma\right)$ and $V_{k}\left(\gamma\right)$ is
given by (31). This completes the proof.\balance


\end{document}